\documentclass[onecolumn,showpacs,preprintnumbers,amsmath,amssymb]{revtex4}

\usepackage{cmap}
\usepackage{bm,amsmath,amssymb,latexsym,mathrsfs,graphicx,enumerate}
\usepackage[mathcal]{euscript}
\usepackage{hyperref}
\usepackage{epsfig}

\usepackage{amsmath}
\usepackage{graphics}
\usepackage{epstopdf}

\usepackage{caption}
\DeclareGraphicsExtensions{.bmp,.png,.jpg}

\begin{document}

\title[Guided Waves in Asymmetric Hyperbolic Slab Waveguides. The TM Mode Case ]
{Guided Waves in Asymmetric Hyperbolic Slab Waveguides. The TM Mode
Case}

\author{Ekaterina I. Lyashko$^{1}$,  Andrey I. Maimistov$^{1,2}$}
\affiliation{\normalsize \noindent $^1$ Department of General and
Applied Physics, Moscow Institute of Physics
and Technology, Dolgoprudny, Moscow region, 141700 Russia \\
$^2$ Department of Solid State Physics and Nanostructures, National
Nuclear Research University
Moscow Engineering Physics Institute, Moscow, 115409  \\
E-mails: ostroukhova.ei@gmail.com, ~~aimaimistov@gmail.com \\
}
\date{\today}

\begin{abstract}
Nonlinear guided wave modes in an asymmetric slab waveguide formed
by an isotropic dielectric layer placed on a linear or nonlinear
substrate and covered by a hyperbolic material are investigated.
Optical axis is normal to the slab plane. The dispersion relations
for TM waves are found. It is shown that there are additional
cut-off frequencies for each TM mode. The effects of the
nonlinearity on the dispersion relations are investigated and
discussed. There are the modes, which are corresponded with
situation where the peak of electric field is localized in the
nonlinear substrate. These modes are absent in the linear waveguide.
To excite these modes the power must exceed certain threshold value.

\end{abstract}


\maketitle

\section{Introduction}

Hyperbolic metamaterials are artificially structured uniaxial
materials with opposite signs of permittivity (or magnetic
permeability) for ordinary and extraordinary waves
\cite{Noginov:09,XNi:11,Drachev:13,Prashant:14}. The hyperbolic
materials can be fabricated as a multilayer structure consisting of
alternating metallic and dielectric 
layers \cite{Wood:06,Kivshar:13,Othman:13}, or as a nanowire
structure consisting of metallic nanorods embedded in a dielectric
host.

Due to the unusual hyperbolic form of the iso-frequency dispersion
surfaces unlimited high values of the wave vector are possible in
such materials. It results in the numerous effects, among which are
the Purcell enhancement of the spontaneous emission rate in
hyperbolic metamaterials
\cite{Poddubny:12,Poddubny:13,Ward:13,Ferrari:14} and the
subwavelength resolution effect \cite{Wood:06,Benedict:13}. The
optical phenomena on the interface between conventional dielectric
and hyperbolic metamaterial has also attracted attention. The
surface waves were studied in \cite{Zapata:13}.  The extremely large
Goos–-H\"{a}nchen shift has been studied in some details in
\cite{JingZhao:13}.

The linear plasmonic planar waveguide cladded by hyperbolic
metamaterials 
was proposed and investigated in
\cite{Kildishev:14,Kildishev:15}. The linear guided waves properties
of such hyperbolic slab waveguide were discussed in
\cite{Lyashko:15}. It was shown that for the TM wave number of
guided modes is limited. Each of these modes have two cut-off
frequencies. One of them corresponds to the mode appearance, another
corresponds to the mode disappearance. There is region of parameters
in which only single mode exists in this waveguide. It is worth
noting that this phenomenon is unavailable in the case of a
conventional waveguide. Usually the number of modes increases with
core thickness or radiation frequency, and only single cut-off
frequency exists for each mode.

The purpose of this paper is to investigate the dispersion
properties of the nonlinear guided waves in a planar asymmetric
waveguide. It is assumed that linear isotropic film (i.e., core of
waveguide) is placed on a nonlinear substrate. The core of waveguide
is covered by the linear hyperbolic material. The anisotropy axis of
the cladding layer is normal to the interface between core and
surroundings. In the planar geometry, as is known
\cite{Hansperger:84}, guided waves can be separated into two gropes
of waves, which have a different polarization. These waves are
referred to as TE and TM waves. In general case the nonlinear
polarization entangles the TE and TM waves. The simplest case where
only TM kind of the wave is exited will be considered.

The electric and magnetic field distributions and the dispersion
relations will be found for the slab waveguide under consideration.
If the extra-ordinary permittivity is less then a permittivity of
core of waveguide, each TM mode of the hyperbolic waveguide both in
the linear and nonlinear cases has an additional cut-off frequency
and a transition point. Both the cut-off frequency and the
transition point depend on power that is transferred by guided wave.
If power exceeds a certain critical value, for each wave mode
appears additional solution branch. The guiding waves, that
correspond to these branches, differ in transverse configuration of
field.

\section{Constituent equations}

Let us consider the uniaxial anisotropic dielectric material. The
vector $\mathbf{a}$ is the unit vector of an optical axis. In this
case the Fourier component of the vector of dielectric displacement
$\mathbf{D}$ can be written as
$$
\mathbf{D} =\varepsilon_o\mathbf{E}+
    (\varepsilon_e-\varepsilon_o)(\mathbf{a}\cdot
    \mathbf{E})\mathbf{a} +4\pi \mathbf{P}_{nl},
$$
where $\mathbf{E} = \mathbf{E}(\mathbf{r},\omega)$ is the Fourier
component of the electric field, $\varepsilon_o$ is an ordinary
permittivity, and $\varepsilon_e$ is an extra-ordinary permittivity.
Theses values are the linear principal dielectric constants for
anisotropic material. Nonlinear properties of the material is
defined by the nonlinear polarization vector $\mathbf{P}_{nl}$.

The Fourier component of the electric field $\mathbf{E}(\omega,
\mathbf{r})$ in a nonmagnetic medium is governed by the following
wave equation
\begin{equation}\label{eq:LM:1}
    \left(\nabla^2+k_0^2\varepsilon_o\right) \mathbf{E} + \frac{\Delta\varepsilon}{\varepsilon_o}
    \left[(\mathbf{a}\cdot \nabla)\nabla +k_0^2\varepsilon_o\mathbf{a} \right](\mathbf{a}\cdot \mathbf{E})
    +4\pi k_0^2\mathbf{P}_{nl} =0,
\end{equation}
where $k_0=\omega/c$, $\Delta\varepsilon = \varepsilon_e -
\varepsilon_o$. In the case of isotropic medium this equation is
reduced to usual wave equation.


Let us consider a slab waveguide. We assume, that the material of
the film (waveguide core) is nonmagnetic $\mu_2 =1$ and has an
isotropic permittivity $\varepsilon_2$. The film thickness is $h$.
The dielectric film is placed on a linear or nonlinear substrate and
covered by the uniaxial hyperbolic metamaterial which is
characterized by symmetric dielectric tensor with the principal
dielectric constants $\varepsilon_o$, and $\varepsilon_e$, and the
magnetic permeability $\mu_1=1$. The anisotropy axis is aligned with
a unit normal vector to the interface, i.e., along the $OX$
direction. 
Axes $OY$ and $OZ$ are parallel to the
interface. Axis $OZ$ is directed along the wave propagation. In this
case the Maxwell equations are invariant under the shifting along
$OY$ axis. Thus, the strengths of electric and magnetic fields of
the guided wave are independent on variable $y$. From it follows,
that the wave equation (\ref{eq:LM:1}) is splitting into two
independent systems of equations describing the TE and TM waves 
\begin{eqnarray}
   && \left(\frac{\varepsilon_e(x)}{\varepsilon_o(x)}\frac{\partial^2}{\partial
   x^2} +\frac{\partial^2}{\partial z^2}
   +k_0^2\varepsilon_e(x)\right)E_x +4\pi k_0^2P_{nl,x} =0, \nonumber\\
   && \left(\frac{\partial^2}{\partial x^2} +\frac{\partial^2}{\partial z^2}
   +k_0^2\varepsilon_o(x)\right)E_y +4\pi k_0^2P_{nl,y}
   =0,\label{eq:LM:2} \\
   && \left(\frac{\partial^2}{\partial x^2} +\frac{\partial^2}{\partial z^2}
   +k_0^2\varepsilon_o(x)\right)E_z +\frac{\Delta\varepsilon(x)}{\varepsilon_o(x)}
   \frac{\partial^2}{\partial x\partial z}E_x +4\pi k_0^2P_{nl,z}
   =0. \nonumber
\end{eqnarray}

The simple model of nonlinear medium will be considered
\cite{Mihalache:86,Agranowich:86}. It what follows we will assumed
that
\begin{equation}\label{eq:LM:assumpt}
   \mathbf{P}_{nl}= \varepsilon_K(|E_y|^2+|E_z|^2)\mathbf{E}
\end{equation}

The TE wave is defined by the following electric and magnetic
vectors $\mathbf{E}=(0, E_y, 0)$ and $\mathbf{H}=(H_x, 0, H_z)$.
Hence, the wave equation for $E_y$ from (\ref{eq:LM:2}) can be used
to describe the nonlinear wave propagation in this slab waveguide.
TE wave is the ordinary wave and the new results in this case are
not expected.

The electric and magnetic vectors of the TM wave are
$\mathbf{E}=(E_x, 0, E_z)$ and $\mathbf{H}=(0, H_y, 0)$.

The components of these vectors are related by the Maxwell
equations, that are

\begin{eqnarray}
  && \qquad \qquad \qquad H_y =\frac{i}{k_0}\left(\frac{\partial E_z}{\partial x}-\frac{\partial E_x}{\partial z}\right),  \\
  && \frac{\partial H_y}{\partial z} = ik_0\varepsilon_e(x)E_x+4\pi ik_0 P_{nl, x},
  \quad \frac{\partial H_y}{\partial x} = -ik_0 \varepsilon_o(x)E_z
  -ik_04\pi P_{nl, z}.
\end{eqnarray}

The linear principal dielectric constants are piecewise functions :
\begin{equation}\label{eq:epsilon:distrib}
\varepsilon_o(x)=\left\{
\begin{array}{ccc}
  \varepsilon_1 & x<0, \\
  \varepsilon_2 & 0 \leq x \leq h, \\
  \varepsilon_o & x>h, \\
\end{array}\right.\quad \varepsilon_e(x)=\left\{
\begin{array}{ccc}
  \varepsilon_1 & x<0, \\
  \varepsilon_2 & 0 \leq x \leq h, \\
  \varepsilon_e & x>h, \\
\end{array}\right.
\end{equation}

\section{Linear slab waveguide}
\subsection{Electromagnetic field distributions}

\noindent As the waveguide is homogeneous medium along the axis
$OZ$, the electric and magnetic vectors of the TM wave can be
represented by the following expressions: $\mathbf{E}=(\tilde{E}_x,
0, \tilde{E}_z)e^{i\beta z}$ and $\mathbf{H}=(0, \tilde{H}_y,
0)e^{i\beta z}$, where $\beta$ is propagation constant
\cite{Hansperger:84}. From the system of equations (\ref{eq:LM:2})
the equations for $\tilde{E}_z$ result in
\begin{eqnarray}
  x<0   &~& \frac{\partial^2}{\partial x^2}\tilde{E}_z+(k_0^2\varepsilon_1 - \beta^2)\tilde{E}_z =0,   \nonumber\\
  0\leq x\leq h &~& \frac{\partial^2}{\partial x^2}\tilde{E}_z+(k_0^2\varepsilon_2 - \beta^2)\tilde{E}_z =0, \label{eq:LM:Lin:5}\\
  x>h   &~& \frac{\partial^2}{\partial x^2}\tilde{E}_z+\frac{\varepsilon_o}{\varepsilon_e}(k_0^2\varepsilon_e - \beta^2)\tilde{E}_z =0,  \nonumber
\end{eqnarray}

The solutions of these equations with taking into account the
boundary conditions at $x\to \pm \infty$ (i.e.,outside of waveguide
all fields are disappeared) $\tilde{E}_z(x)$ and magnetic field
distributions $\tilde{H}_y(x)$ are
\begin{eqnarray}
  x<0   &~& \tilde{E}^{(1)}_z(x) = A_0e^{px}, \qquad \qquad\qquad \tilde{H}^{(1)}_y = -\frac{ik_0\varepsilon_1}{p}A_0e^{px}, \label{eq:LM:Lin:6a}\\
  0\leq x\leq h &~& \tilde{E}^{(2)}_z(x) = B_1e^{i\kappa x}+B_2e^{-i\kappa x},\quad
   \tilde{H}^{(2)}_y(x) = -\frac{k_0\varepsilon_2}{\kappa}\left(B_1e^{i\kappa x}-B_2e^{-i\kappa x}\right),\label{eq:LM:Lin:6b}\\
  x>h   &~& \tilde{E}^{(3)}_z(x) = Ce^{-qx}, \qquad\qquad\qquad \tilde{H}^{(3)}_y(x) = \frac{ik_0\varepsilon_o}{q}
    Ce^{-qx},\label{eq:LM:Lin:6c}
\end{eqnarray}
where the real positive value
$$
p=\sqrt{\beta^2-k_0^2\varepsilon_1}, \quad
q=\sqrt{\varepsilon_o/\varepsilon_e(\beta^2-k_0^2\varepsilon_e}),
\quad \kappa=\sqrt{k_0^2\varepsilon_2-\beta^2},
$$
are introduced. If the $\kappa$ is imaginary value then the
solutions (\ref{eq:LM:Lin:6a})--(\ref{eq:LM:Lin:6c}) describe the
coupled surface wave propagation.

The electromagnetic field is confined in waveguide if the conditions
$$k_0^2\varepsilon_1<\beta^2 < k_0^2\varepsilon_2, \quad
(\varepsilon_o/\varepsilon_e)(\beta^2-k_0^2\varepsilon_e)>0 $$ are
held. In the case of hyperbolic material with $\varepsilon_o>0$,
$\varepsilon_e<0$ the second inequality breaks down for any $\beta$.
The evanescent wave comes into covering layer. Oppositely, in the
case of hyperbolic material with $\varepsilon_o <0$,
$\varepsilon_e>0$ the second inequality can be held, if $\beta^2
<k_0^2\varepsilon_e$. Let us introduce the effective index $n_{eff}
=\beta/k_0$. Thus the confinement condition reads as
\begin{equation}\label{eq:LM:Lin:7}
    \varepsilon_1<n_{eff}^2< \min(\varepsilon_e,~\varepsilon_2).
\end{equation}

The continuity conditions for tangent components of both electric
and magnetic field vectors at $x=0$ and $x=h$ result in following
relations:
\begin{eqnarray}
   && A_0=B_1+B_2, \qquad
     \frac{i\varepsilon_1}{p}A_0= \frac{\varepsilon_2}{\kappa}(B_1-B_2), \nonumber\\
   && Ce^{-qh} = B_1e^{i\kappa h}+ B_2e^{-i\kappa h}, \label{eq:LM:Lin:continu:1}\\
   && \frac{i\varepsilon_o}{q}Ce^{-qh}
   =-\frac{\varepsilon_2}{\kappa}\left( B_1e^{i\kappa h}- B_2e^{-i\kappa
   h}\right).\nonumber
\end{eqnarray}
From the first and second equations of this system of the linear
equations follows that
$$
B_1=\frac{A_0}{2}\left(1+ i\frac{\varepsilon_1\kappa}{\varepsilon_2
p}\right), \quad B_2=\frac{A_0}{2}\left(1-
i\frac{\varepsilon_1\kappa}{\varepsilon_2 p}\right).
$$
Amplitude $C$ is
$$
C= \frac{1}{2}\left[\left(1+
i\frac{\varepsilon_1\kappa}{\varepsilon_2 p}\right)e^{i\kappa h} +
\left(1-i\frac{\varepsilon_1\kappa}{\varepsilon_2
p}\right)e^{-i\kappa h} \right]A_0 e^{qh}.
$$
Thus the electric field distribution of the TM wave can be written
as
\begin{eqnarray}
  x<0   &~& \tilde{E}^{(1)}_z(x) = A_0e^{px},  \nonumber\\
  0\leq x\leq h &~& \tilde{E}^{(2)}_z(x) = A_0 \left(\cos(\kappa x) - \frac{\varepsilon_1\kappa}
  {\varepsilon_2p}\sin(\kappa x)\right),\label{eq:LM:Lin:6}\\
  x>h   &~& \tilde{E}^{(3)}_z(x) = A_0\left(\cos(\kappa h) - \frac{\varepsilon_1\kappa}
  {\varepsilon_2p}\sin(\kappa h)\right)e^{-q(x-h)},\nonumber
\end{eqnarray}

\subsection{Dispersion relation}

The dispersion relation for the localized in the waveguide
electromagnetic waves represents the relationship between a
propagation constant $\beta$ and a frequency $\omega$. The
dispersion relations for the guided waves under consideration can be
determined using the continuity conditions of an electric and
magnetic field on interferences, as was done in \cite{Lyashko:15}.
The linear system of algebraic equations (\ref{eq:LM:Lin:continu:1})
has non-zero solutions if the determinant of this system is zero.
This claim results in the desired dispersion relation. As a
consequence the following expression can be found
\begin{equation}\label{eq:LM:11}
    e^{2i\kappa h}\left(\frac{1-i\xi_p}{1 +i\xi_p}\right)
\left(\frac{1+i\xi_q}{1-i\xi_q}\right) =1,
\end{equation}
where following parameters
$$
\xi_p=\frac{\varepsilon_2 p}{\varepsilon_1 \kappa}, \quad
\xi_q=\frac{\varepsilon_2 q}{|\varepsilon_o| \kappa}
$$
are used. The Goos-–H\"{a}nchen phase shifts $\phi_{q}$ and
$\phi_{p}$ are defined by following expressions $\xi_q =
-\tan(\phi_{q}/2)$ and $\xi_p = \tan(\phi_{p}/2)$. Using these
expressions we can write the dispersion relation in form
\begin{equation}\label{eq:LM:12}
    2\kappa h - \phi_p-\phi_q = 2\pi m, \quad m = 0,~1,~2,\ldots
\end{equation}
If the effective index of refraction $n_{eff}$ is used, than equation
(\ref{eq:LM:12}) can be written as
\begin{equation}\label{eq:LM:13}
k_0h\sqrt{\varepsilon_2-n_{eff}^2} =
\arctan\left(\frac{\varepsilon_2}{\varepsilon_1}\sqrt{\frac{n_{eff}^2-\varepsilon_1}{\varepsilon_2-n_{eff}^2}}\right)
-\arctan\left(\frac{\varepsilon_2}{\sqrt{\varepsilon_e|\varepsilon_o|}}
\sqrt{\frac{\varepsilon_e -n_{eff}^2}{\varepsilon_2-n_{eff}^2}}\right)
+\pi m.
\end{equation}
Thus the dispersion relation as the implicit function $n_{eff}$ of
the $\varepsilon_1$, $\varepsilon_2$, $\varepsilon_e$, $h$, $\omega
= ck_0$ is obtained.

\begin{figure}[h!]
    \center{\includegraphics[scale=0.90]{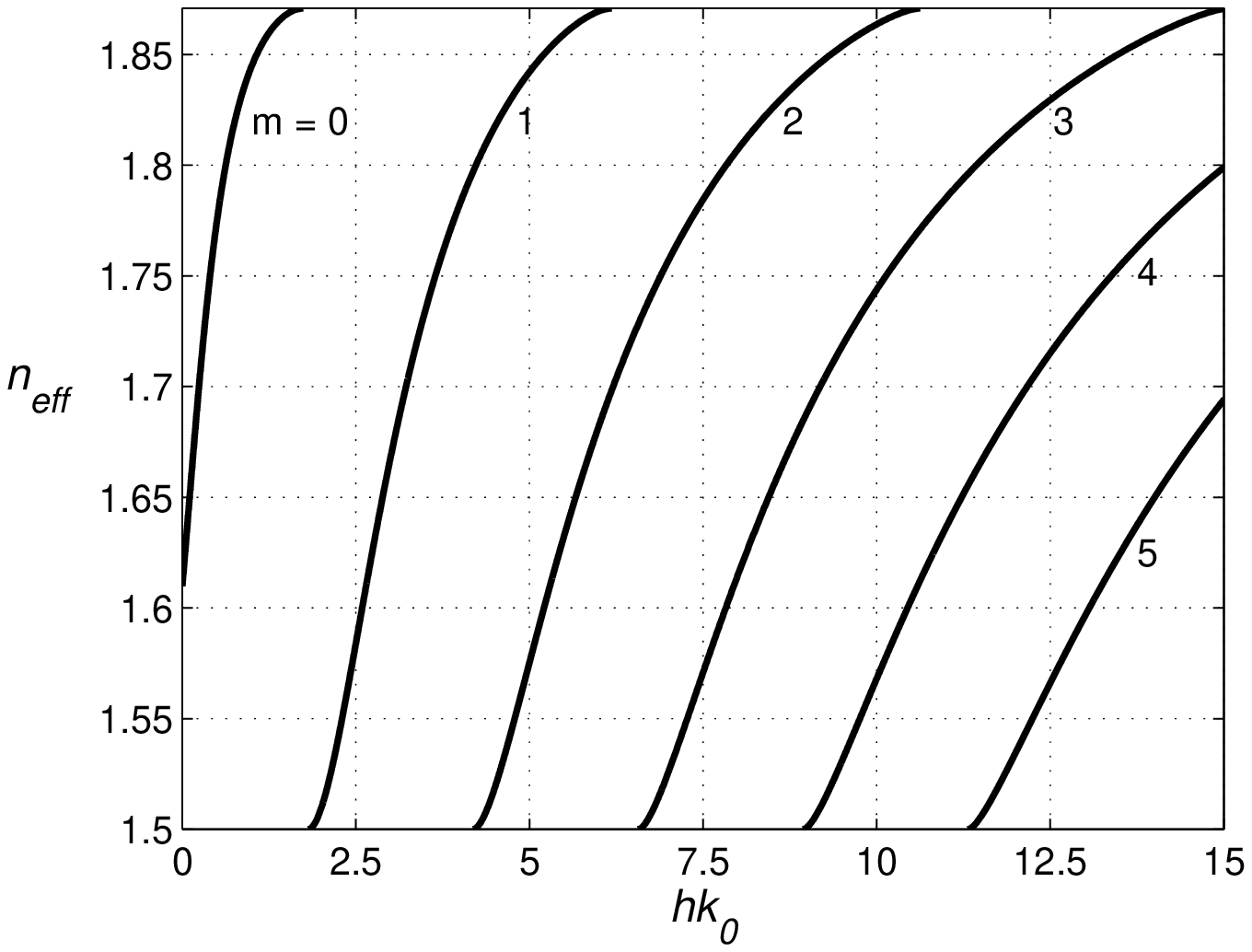}}
    \caption{Dispersion curves for TM modes of the hyperbolic waveguide. $\varepsilon_e<\varepsilon_2$ . }
    \label{LM:HLH:2}
\end{figure}

Let it be $\varepsilon_e<\varepsilon_2$. In this case the two
cut-off frequencies series take place. At $n_{eff}^2 =
\varepsilon_1$ they are $V_{cm}^{(1)}$, and at $n_{eff}^2 =
\varepsilon_e$ they are $V_{cm}^{(2)}$, where
\begin{eqnarray}
  V_{cm}^{(1)}=k_0h\sqrt{\varepsilon_2-\varepsilon_1} &=& \pi m -
  \arctan\left(\frac{\varepsilon_2}{\sqrt{\varepsilon_e|\varepsilon_o|}}
\sqrt{\frac{\varepsilon_e -\varepsilon_1}{\varepsilon_2-\varepsilon_1}}\right), \label{eq:LM:cutoff:1}\\
  V_{cm}^{(2)}=k_0h\sqrt{\varepsilon_2-\varepsilon_e}  &=& \pi m +
  \arctan\left(\frac{\varepsilon_2}{\varepsilon_1}\sqrt{\frac{\varepsilon_e-\varepsilon_1}{\varepsilon_2-\varepsilon_e}}\right).
  \label{eq:LM:cutoff:2}
\end{eqnarray}
As the wavenumber $k_0$ depends on the frequency of radiation, the
waveguide guides every TM mode only in certain frequency range.

It should be pointed that $V_{c 0}^{(1)}$ is negative quantity, that
is impossible. Hence, the dispersion curve with $m=0$ presents the
special situation. At $m=0$ the dispersion relation is
$$
k_0h\sqrt{\varepsilon_2-n_{eff}^2} =
\arctan\left(\frac{\varepsilon_2}{\varepsilon_1}\sqrt{\frac{n_{eff}^2-\varepsilon_1}{\varepsilon_2-n_{eff}^2}}\right)
-\arctan\left(\frac{\varepsilon_2}{\sqrt{\varepsilon_e|\varepsilon_o|}}
\sqrt{\frac{\varepsilon_e-n_{eff}^2}{\varepsilon_2-n_{eff}^2}}\right).
$$
As the left-hand side of this expression is positive one, then on
the right-hand side the sum of terms is positive too. It results in
$$
\frac{n_{eff}^2-\varepsilon_1}{\varepsilon_1^2} \geq
\frac{\varepsilon_e-n_{eff}^2}{\varepsilon_e|\varepsilon_o|}.$$
Hence, the minimal effective index of this dispersion curve with
$m=0$ is determined by equation
$$
\min n_{eff}^2 = \varepsilon_1+\frac{\varepsilon_1^2(\varepsilon_e
-\varepsilon_1)}{\varepsilon_1^2+\varepsilon_e|\varepsilon_o|}.
$$
It should be remarked that this magnitude of effective index is
achieved at the zero film thickness. In the case of a conventional
asymmetrical slab waveguide the cut-off frequencies corresponding
$m=0$ is positive quantity. The property of the guided mode with
$m=0$ discussed above is the feature of hyperbolic asymmetrical slab
waveguide.

If the zigzag-ray model for light propagation in slab waveguide
\cite{Kogelnik:Weber:74} is used, then effective index can be
represented by the formula $n_{eff} = \sqrt{\varepsilon_2}\sin
\alpha$, where $\alpha$ is the incident angle on the film-substrate
and film-cover interfaces. The inequality (\ref{eq:LM:Lin:7})
results in
\begin{equation}\label{eq:LM:Lin:8}
    \frac{\varepsilon_1}{\varepsilon_2} <\sin^2\alpha <\min\left(\frac{\varepsilon_e}{\varepsilon_2}~,1 \right).
\end{equation}

When $\varepsilon_2 <\varepsilon_e$ inequality (\ref{eq:LM:Lin:8})
is consistent with the usual condition for dielectric waveguides.
The light ray is totally reflected at the film-substrate interface
if $\alpha > \alpha_1=\arcsin(\varepsilon_1/\varepsilon_2)^{1/2}$.
If $\varepsilon_2 >\varepsilon_e$, the second critical angle occurs.
Now the confinement condition reads as $\alpha_1 <\alpha< \alpha_2$,
where $\alpha_2=\arcsin(\varepsilon_e/\varepsilon_2)^{1/2}$. It
means that the two cut-off frequencies take place in the hyperbolic
waveguide under consideration. Fig.\ref{LM:HLH:2} shows the branches
of dispersion curves ($m=0 \div 5$) on condition that
$\varepsilon_1=2.25$, $\varepsilon_2=4.0$, $\varepsilon_e=3.5$, and
$\varepsilon_o=-3.85$.

Now let it be $\varepsilon_e >\varepsilon_2$. In this case only one
cut-off frequencies series takes place:  $V_{c m}^{(1)}$, $m\geq 1$,
which is determined by equation (\ref{eq:LM:cutoff:1}). At $m=0$ guided
mode has the properties identical to that for the case of
$\varepsilon_e <\varepsilon_2$.

\begin{figure}[h!]
    \center{\includegraphics[scale=0.90]{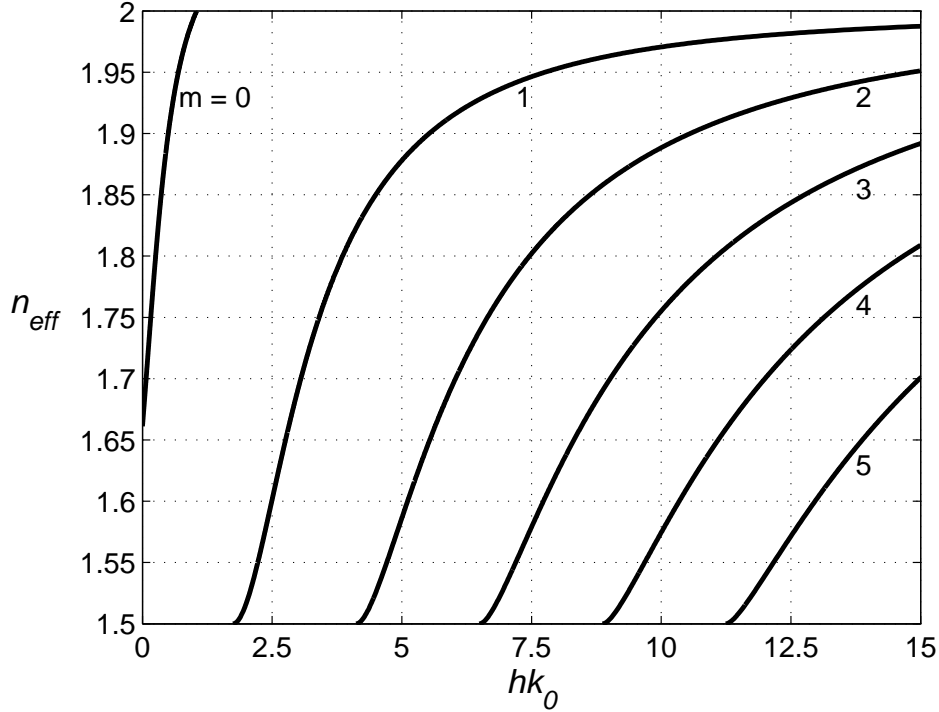}}
    \caption{Dispersion curves for TM modes of the hyperbolic waveguide. $\varepsilon_e<\varepsilon_2$ . }
\label{LM:HLH:3}
\end{figure}

The dispersion curves at $m=0 \div 5$ on condition that
$\varepsilon_1=2.25$, $\varepsilon_2=4.0$, $\varepsilon_e=4.5$, and
$\varepsilon_o=-3.85$ are shown in Fig.\ref{LM:HLH:3}. One can see
that the cut-off frequencies $V_{c m}^{(2)}$ with $m\geq 1$ are
lacking. However mode with $m=0$ has the second cut-off frequency.

\section{Nonlinear waveguide}

\noindent Now the effect of nonlinearity will be considered.
Nonlinear property of the substrate will be described using the
(\ref{eq:LM:assumpt}) \cite{Agranowich:86}. As the electric field
vector of the considered TM waves is $\mathbf{E}=(E_x, 0, E_z)$, the
nonlinear polarization $P_{z,nl}$ can be written as
\begin{equation}\label{eq:LM:assumpt:2}
   P_{z,nl}= \varepsilon_K|E_z|^2E_z.
\end{equation}
Electric field distributions are governed by following equations
\begin{eqnarray}
  x<0   &~& \frac{\partial^2}{\partial x^2}\tilde{E}_z+(k_0^2\varepsilon_1 - \beta^2)\tilde{E}_z +k_0^2\varepsilon_K \tilde{E}_z^3 =0,   \nonumber\\
  0\leq x\leq h &~& \frac{\partial^2}{\partial x^2}\tilde{E}_z+(k_0^2\varepsilon_2 - \beta^2)\tilde{E}_z =0, \label{eq:LM:Nonin:15}\\
  x>h   &~& \frac{\partial^2}{\partial x^2}\tilde{E}_z+\frac{\varepsilon_o}{\varepsilon_e}(k_0^2\varepsilon_e - \beta^2)\tilde{E}_z =0,  \nonumber
\end{eqnarray}
The normal component of the electric field and the tangent component
of magnetic field distribution can be obtained from the following
expression
$$
\tilde{H}_y(x) = \frac{ik_0\varepsilon_e(x)}{k_0^2\varepsilon_e(x)
-\beta^2}\frac{\partial \tilde{E}_z}{\partial x}, \qquad
\tilde{E}_x(x) = \frac{i\beta}{k_0^2\varepsilon_e(x)
-\beta^2}\frac{\partial \tilde{E}_z}{\partial x}.
$$
The permittivity function $\varepsilon_e(x)$ is defined by the
expression (\ref{eq:epsilon:distrib}).

\subsection{Electromagnetic field distributions}

As above the the real positive parameters $
p=\sqrt{\beta^2-k_0^2\varepsilon_1}$, \quad
$q=\sqrt{\varepsilon_o/\varepsilon_e(\beta^2-k_0^2\varepsilon_e})$,
and $ \kappa=\sqrt{k_0^2\varepsilon_2-\beta^2} $ are introduced. The
electric fields in dielectric film and hyperbolic cover have already
been found, eqs. (\ref{eq:LM:Lin:6b}) and (\ref{eq:LM:Lin:6c}).
Electric field distribution in nonlinear substrate is governed by
the first equation of system (\ref{eq:LM:Nonin:15}). Solution of
this equation can be written as
\begin{eqnarray}
  \tilde{E}_z^{(1)} &=&  \sqrt{\frac{2p^2}{k_0^2\varepsilon_K}}\frac{1}{\cosh[p(x-x_0)]},
  \quad \varepsilon_K >0, \label{eq:LM:16a}\\
  \tilde{E}_z^{(1)}&=& \sqrt{\frac{2p^2}{k_0^2|\varepsilon_K|}}\frac{1}{\sinh[p(x-x_0)]},
  \quad \varepsilon_K <0, \label{eq:LM:16b}
\end{eqnarray}
The parameter $x_0$ dictates the location of the electric field
maximum.  In the case of $\varepsilon_K <0$ the function in
(\ref{eq:LM:16b}) tends to infinity at $x_0>0$. So only the case
$x_0>0$ when special point lyes outside of nonlinear medium is
possible. Hence the electric field maximum is located at $x=0$. In
the case of $\varepsilon_K >0$ one can find that $x_0<0$ (the
electric field maximum is located in nonlinear substrate) or $x_0>0$
(the electric field maximum is located at $x=0$). Furthermore we
will consider only the case of $\varepsilon_K >0$.

\begin{figure}[h]
\begin{minipage}[h]{.49\linewidth}
    \centering
    \includegraphics[width=\linewidth]{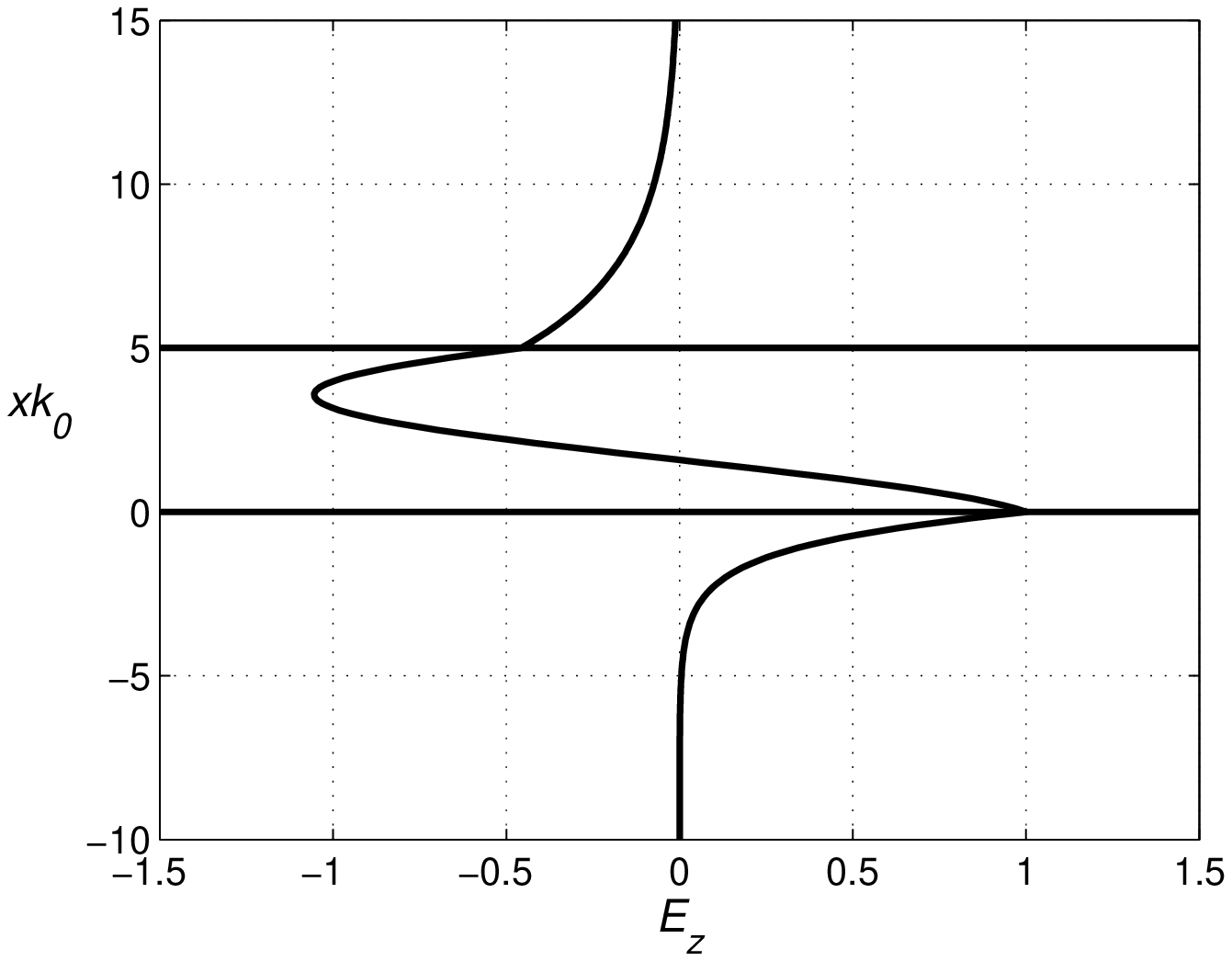}
    \\(a)
\end{minipage}
\hfill
\begin{minipage}[h]{.49\linewidth}
    \centering
    \includegraphics[width=\linewidth]{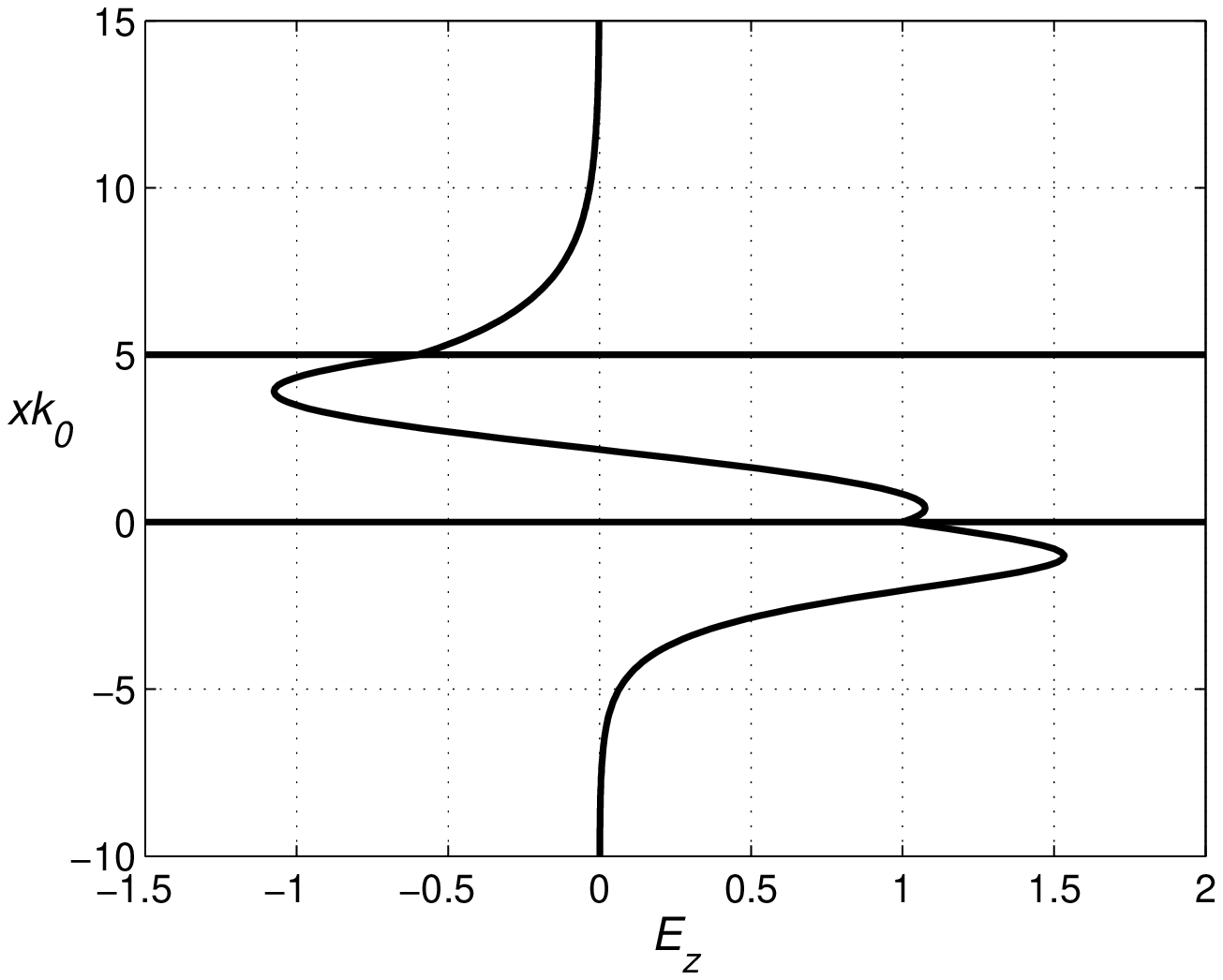}
    \\(b)
\end{minipage}
    \caption{Transverse distribution of the electric field $\tilde{E}_z(x)$ at $m=1$.
    (a) $\nu >0$, (b) $\nu <0$. }
\label{LM:HLH:4}
\end{figure}

The useful parameters are the electric field amplitude at the
boundary of nonlinear substrate $\tilde{A}_0=\tilde{E}_z^{(1)}(x=0)$
and the measure of nonlinearity $\nu$:
$$
\tilde{A}_0 =
\sqrt{\frac{2p^2}{k_0^2\varepsilon_K}}\frac{1}{\cosh(px_0)}=\frac{A_m}{\cosh(px_0)},\quad
\nu = \tanh(px_0) = \mathrm{sgn}(x_0)
\left(1-\frac{k_0^2\varepsilon_K}{2p^2}\tilde{A}_0^2\right)^{1/2}.
$$
It should be pointed that $\tilde{A}_0$ is limited:
$\tilde{A}_0^2\leq 2p^2/k_0^2\varepsilon_K$. $A_m$ is the peak of
the electric field in substrate.  The positive sign of $\nu$
correlates with monotonic behavior of the electric field, the
negative sign of $\nu$ correlates with the localization of the
electric field maximum in nonlinear substrate (see Fig.
\ref{LM:HLH:4}).

With these parameters the expression (\ref{eq:LM:16a}) can be
rewritten as
\begin{equation}\label{eq:LM:18a}
    \tilde{E}_z^{(1)}(x)=\frac{A_m}{\cosh[(px-x_0)]}=\frac{\tilde{A}_0}{\cosh(px)-\nu \sinh(px)}.
\end{equation}
The function describing the electric fields in linear media
$\tilde{E}_z^{(2)}(x)$ and $\tilde{E}_z^{(3)}(x)$ were found above.

The magnetic field distributions are
\begin{eqnarray}
  x<0   &~& \tilde{H}^{(1)}_y(x) = \frac{ik_0\varepsilon_1 A_m\tanh[p(x-x_0)]}{p\cosh[p(x-x_0)]},  \nonumber\\
  0\leq x\leq h &~& \tilde{H}^{(2)}_y(x) = -\frac{k_0\varepsilon_2}{\kappa} \left(B_1e^{i\kappa x} - B_2e^{-i\kappa x}\right),\label{eq:LM:nonlin:H:18}\\
  x>h   &~& \tilde{H}^{(3)}_y(x) = \frac{ik_0\varepsilon_o}{q}Ce^{-qx}.\nonumber
\end{eqnarray}

The continuity conditions for tangent components of the electric and
the magnetic field vectors at $x=0$ and $x=h$ involve the following
relations
\begin{eqnarray}
   && \tilde{A}_0=B_1+B_2, \qquad
     \frac{i\varepsilon_1}{\tilde{p}}\tilde{A}_0= \frac{\varepsilon_2}{\kappa}(B_1-B_2), \label{eq:LM:nonLin:continu:1}\\
   && Ce^{-qh} = B_1e^{i\kappa h}+ B_2e^{-i\kappa h}, \label{eq:LM:nonLin:continu:2}\\
   && \frac{\varepsilon_o}{iq}Ce^{-qh}
   =\frac{\varepsilon_2}{\kappa}\left( B_1e^{i\kappa h}- B_2e^{-i\kappa
   h}\right),\label{eq:LM:nonLin:continu:3}
\end{eqnarray}
where $\tilde{p} = p/\nu$.

The equations (\ref{eq:LM:nonLin:continu:1}) and
(\ref{eq:LM:nonLin:continu:2}) allow one to find the electric field
distributions
\begin{eqnarray}
  x<0   &~& \tilde{E}^{(1)}_z(x) = \frac{\tilde{A}_0\cosh(px_0)}{\cosh[p(x-x_0)]},  \nonumber\\
  0\leq x\leq h &~& \tilde{E}^{(2)}_z(x) = \tilde{A}_0 \left(\cos(\kappa x) - \nu\frac{\varepsilon_1\kappa}
  {\varepsilon_2p}\sin(\kappa x)\right),\label{eq:LM:nonlin:18}\\
  x>h   &~& \tilde{E}^{(3)}_z(x) = \tilde{A}_0\left(\cos(\kappa h) - \nu\frac{\varepsilon_1\kappa}
  {\varepsilon_2p}\sin(\kappa h)\right)e^{-q(x-h)}.\nonumber
\end{eqnarray}
If $\nu \to 1$, then the results of the linear waveguide are
reproduced.

\subsection{Dispersion relation}

The dispersion relation for the guided TM wave in the slab waveguide
under consideration results from the condition that the determinant
of the system of linear equations (\ref{eq:LM:nonLin:continu:1})--
(\ref{eq:LM:nonLin:continu:3}) is equal to zero. It leads one to the
following expression
\begin{equation}\label{eq:LM:nonlin:disper}
k_0h\sqrt{\varepsilon_2-n_{eff}^2} =
\arctan\left(\frac{\varepsilon_2}{\nu\varepsilon_1}\sqrt{\frac{n_{eff}^2-\varepsilon_1}{\varepsilon_2-n_{eff}^2}}\right)
-\arctan\left(\frac{\varepsilon_2}{\sqrt{\varepsilon_e|\varepsilon_o|}}
\sqrt{\frac{\varepsilon_e -n_{eff}^2}{\varepsilon_2-n_{eff}^2}}\right)
+\pi m,
\end{equation}
where the effective index $n_{eff}$ is used. By using the definition
of parameter $\nu$ this expression can be rewritten as
\begin{equation}\label{eq:LM:nonlin:disper:2}
k_0h\sqrt{\varepsilon_2-n_{eff}^2} =\mathrm{sgn}(\nu)
\arctan\left(\frac{\varepsilon_2}{\varepsilon_1}
\sqrt{\frac{(n_{eff}^2-\varepsilon_1)^2}{(\varepsilon_2-n_{eff}^2)(n_{eff}^2-\varepsilon_{nl})}}\right)
-\arctan\left(\frac{\varepsilon_2}{\sqrt{\varepsilon_e|\varepsilon_o|}}
\sqrt{\frac{\varepsilon_e -n_{eff}^2}{\varepsilon_2-n_{eff}^2}}\right)
+\pi m,
\end{equation}
where $\varepsilon_{nl}=\varepsilon_1+\varepsilon_K\tilde{A}_0^2/2$.
$\mathrm{sgn}(\nu)$ corresponds \textbf{to} two branches of the dispersion
relation. These $\nu$-branches each contain the series of the curves
marked by integer $m$. At positive $\nu$ the expression
(\ref{eq:LM:nonlin:disper:2}) is reduced to the dispersion relation for
linear guided waves (\ref{eq:LM:13}) when $\nu \to 1$ or
$\varepsilon_K \to 0$. The negative $\nu$ corresponds with situation
where the peak of the electric field is localized in the nonlinear
substrate. That is impossible in the linear waveguide case
\cite{Stegeman:Seaton:85,Mihalache:86,Lederer:83b}.

Dispersion relation (\ref{eq:LM:nonlin:disper:2}) involves the
inequality for effective index
\begin{equation}\label{eq:LM:nonlin:20}
    \varepsilon_{nl} \leq n_{eff}^2 <\min(\varepsilon_e,~\varepsilon_2).
\end{equation}
Comparison of the inequality (\ref{eq:LM:Lin:7}) with
(\ref{eq:LM:nonlin:20}) under taking into account definition of
$\varepsilon_{nl}$ shows decreasing of the interval of relevant
effective indexes.

At $n_{eff}^2 = \varepsilon_{nl}$ expression
(\ref{eq:LM:nonlin:disper:2}) results in
\begin{equation}\label{eq:LM:nonlin:transient}
    k_0h\sqrt{\varepsilon_2-\varepsilon_{nl}} =\pi \left(m\pm
\frac{1}{2}\right)
-\arctan\left(\frac{\varepsilon_2}{\sqrt{\varepsilon_e|\varepsilon_o|}}
\sqrt{\frac{\varepsilon_e
-\varepsilon_{nl}}{\varepsilon_2-\varepsilon_{nl}}}\right).
\end{equation}
Hence, the point $n_{eff}=\varepsilon_{nl}^{1/2}$ is the common one
for both dispersion curve marked $m$ and dispersion curve with
$m+1$. Therewith the dispersion curve marked $m$ belongs to positive
$\nu$ branch, but the mode marked $m+1$ belongs to negative $\nu$
branch. Thus the points $n_{eff}=\varepsilon_{nl}^{1/2}$ connect the
two $\nu$-branches of the dispersion curves. These points are
referred to as the transition points \cite{Lederer:85}.

\begin{figure}[h]
\begin{minipage}[h]{.49\linewidth}
    \centering
    \includegraphics[width=\linewidth]{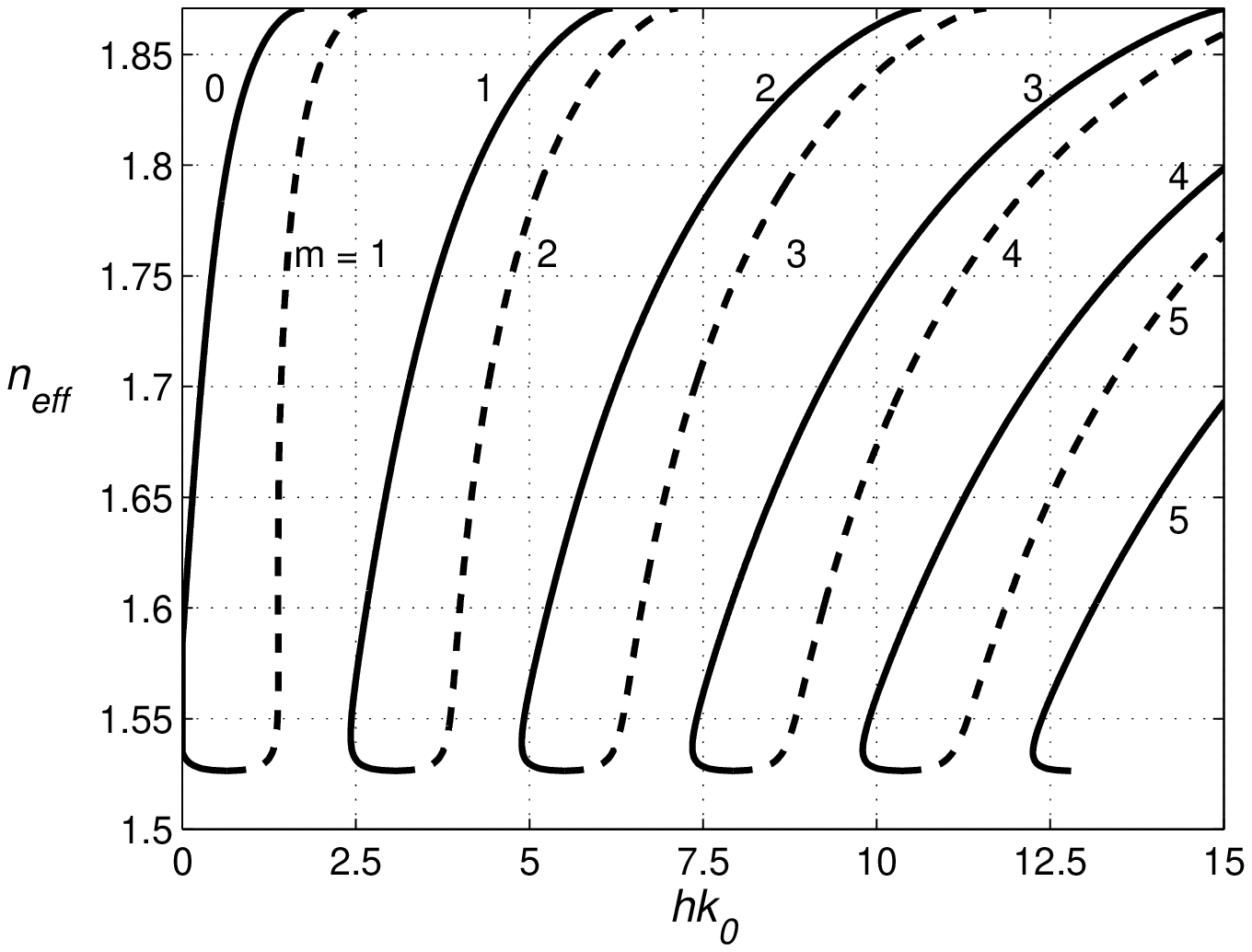}
    \\(a)
\end{minipage}
\hfill
\begin{minipage}[h]{.49\linewidth}
     \centering
     \includegraphics[width=\linewidth]{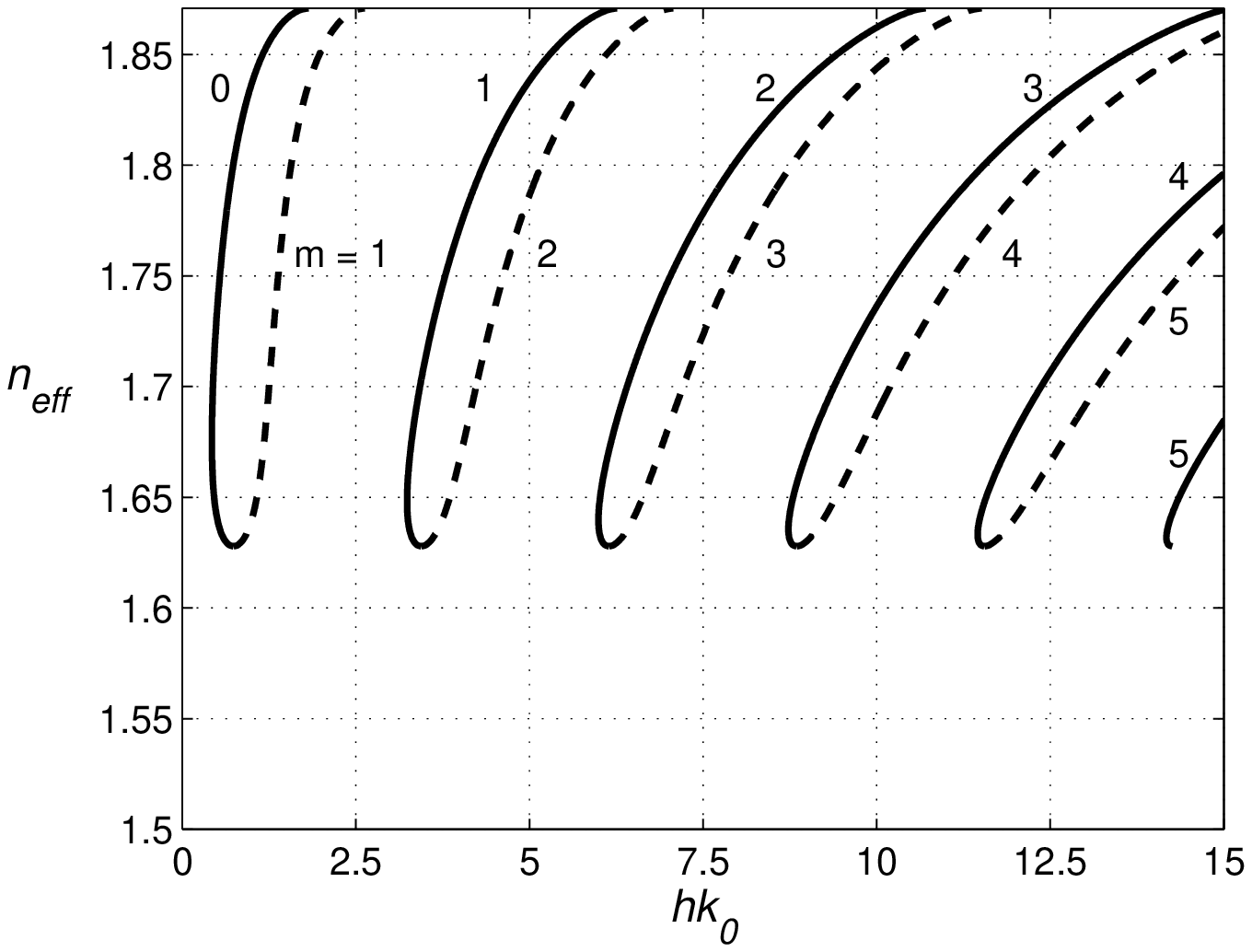}
     \\(b)
\end{minipage}
    \caption{Dispersion curves for different value of $\varepsilon_{nl}$ and mode mark $m$.
    (a) $\varepsilon_{nl}=2.33$, (b) $\varepsilon_{nl}=2.65$. }
\label{LM:HLH:5}
\end{figure}

The case of $\varepsilon_2>\varepsilon_e$ is considered. Fig.
\ref{LM:HLH:5} shows the dispersion curves for different
$\varepsilon_{nl}$ (i.e., for different values of $\tilde{A}_0^2$)
at $\varepsilon_1=2.25$, $\varepsilon_2=4$, $\varepsilon_e=3.5$,
$\varepsilon_0=-3.85$. Mode mark $m$ ranges from 0 to 5. Solid
(dashed) curves correspond to $\nu>0$ ($\nu<0$). As the $k_0h$
increases, the electric field distribution is transformed, so that
the electric field maximum is beginning to shift from film to
substrate and it crosses the interface $x=0$ at transient point.
It should be pointed that positive value of $x_0$ is the position of
the virtual peak. Thus the real electric field maximum is located at
$x=0$ as long as $k_0h$ does not achieve the transient point.

If $\varepsilon_2>\varepsilon_e$ then except for transition points
(\ref{eq:LM:nonlin:transient}) two series of the cut-off frequencies
exists:
\begin{equation}\label{eq:LM:nonlin:cutoff:1}
    V_{cm}^{\pm}=k_0h\sqrt{\varepsilon_2-\varepsilon_e} =\pm
    \arctan\left(\frac{\varepsilon_2}{\varepsilon_1}
\sqrt{\frac{(\varepsilon_e-\varepsilon_1)^2}{(\varepsilon_2
-\varepsilon_e)(\varepsilon_e-\varepsilon_{nl})}}\right)+\pi m.
\end{equation}

As in the case of the linear waveguide here the dispersion curves marked
by $m=0$ represent the special case. If $\mathrm{sgn}(\nu)=-1$ then
the curve with $m=0$ is absent. For $\mathrm{sgn}(\nu)=+1$ the
cut-off frequency formally is negative. However, it means that
minimum of the effective index is more then $\varepsilon_{nl}$ as
previously. Minimum of the effective index can be written as $\min
n_{eff}^2 =\varepsilon_{nl}+ \vartheta $, where $\vartheta$ is
minimal root of the equation
$$
\varepsilon_e|\varepsilon_o|(\vartheta+\varepsilon_{nl}-\varepsilon_1)^2=
\varepsilon_1^2\vartheta(\varepsilon_e-\varepsilon_1-\vartheta),
$$

\begin{figure}[h]
    \center{\includegraphics[scale=0.8]{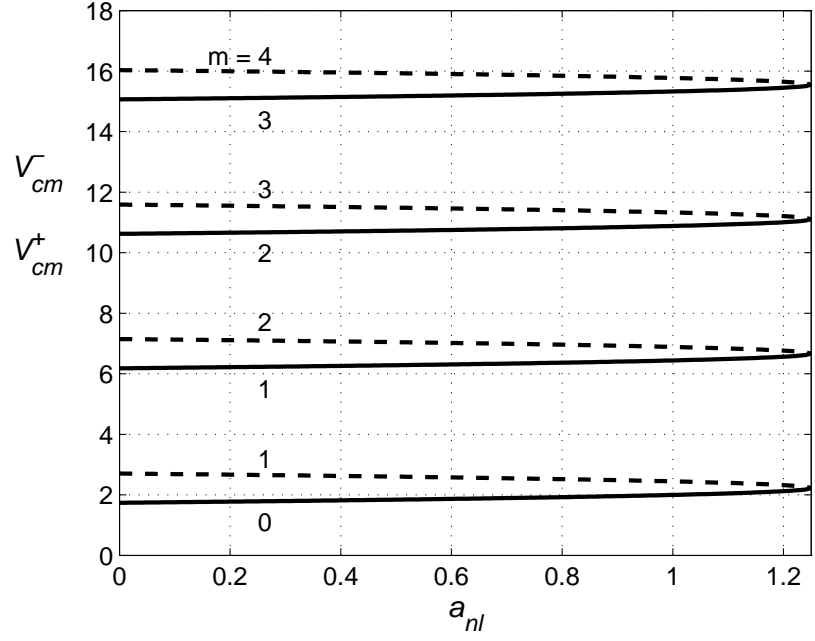}}
    \caption{ Cut-off frequencies $V_{cm}^{\pm}/\sqrt{\varepsilon_2-\varepsilon_e}$ as function of
    intensity parameter $a_{nl}$ at some mode mark $m$.
    }
\label{LM:HLH:6}
\end{figure}

Let us denote intensity parameter $a_{nl}$ as difference
$\varepsilon_{nl}-\varepsilon_1$. Fig. \ref{LM:HLH:6} presents the
cut-off frequencies $V_{cm}^{\pm}$ on the intensity parameter. Solid
(dashed) curve shows behavior of
$V_{cm}^{+}$ ($V_{cm}^{-}$) for $\nu>0 (\nu <0$). On this figure
$k_0h$ corresponding with the cut-off frequencies are indicated. In
the limit $a_{nl} = p^2/k_0^2$, where $\nu=0$, curves of the
different $\nu$-branch converge to a common point. This convergence
therewith takes place for the curves with ($\nu>0$, $m$) and
($\nu<0$, $m+1$).

\begin{figure}[h]
    \center{\includegraphics[scale=0.8]{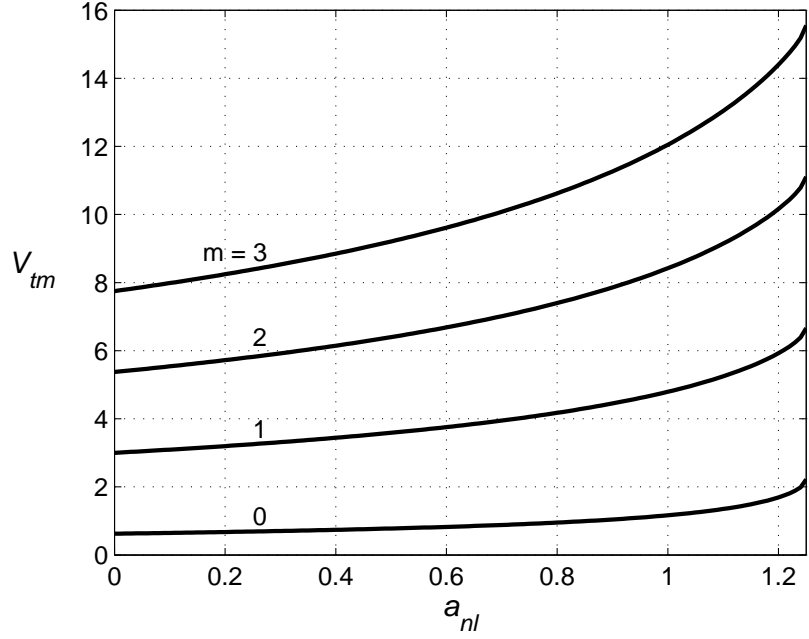}}
    \caption{Dependence of the transition point position on intensity parameter $a_{nl}$
     at some mode mark $m$.
     }
\label{LM:HLH:7}
\end{figure}

The curves in the Fig. \ref{LM:HLH:6} demonstrate a weak dependence
of the cut-off frequencies on intensity parameter. The frequency
corresponding to a transition point essentially depends on the
intensity parameter (Fig.\ref{LM:HLH:7}).

\begin{figure}[h]
\begin{minipage}[h]{.49\linewidth}
    \centering
    \includegraphics[width=\linewidth]{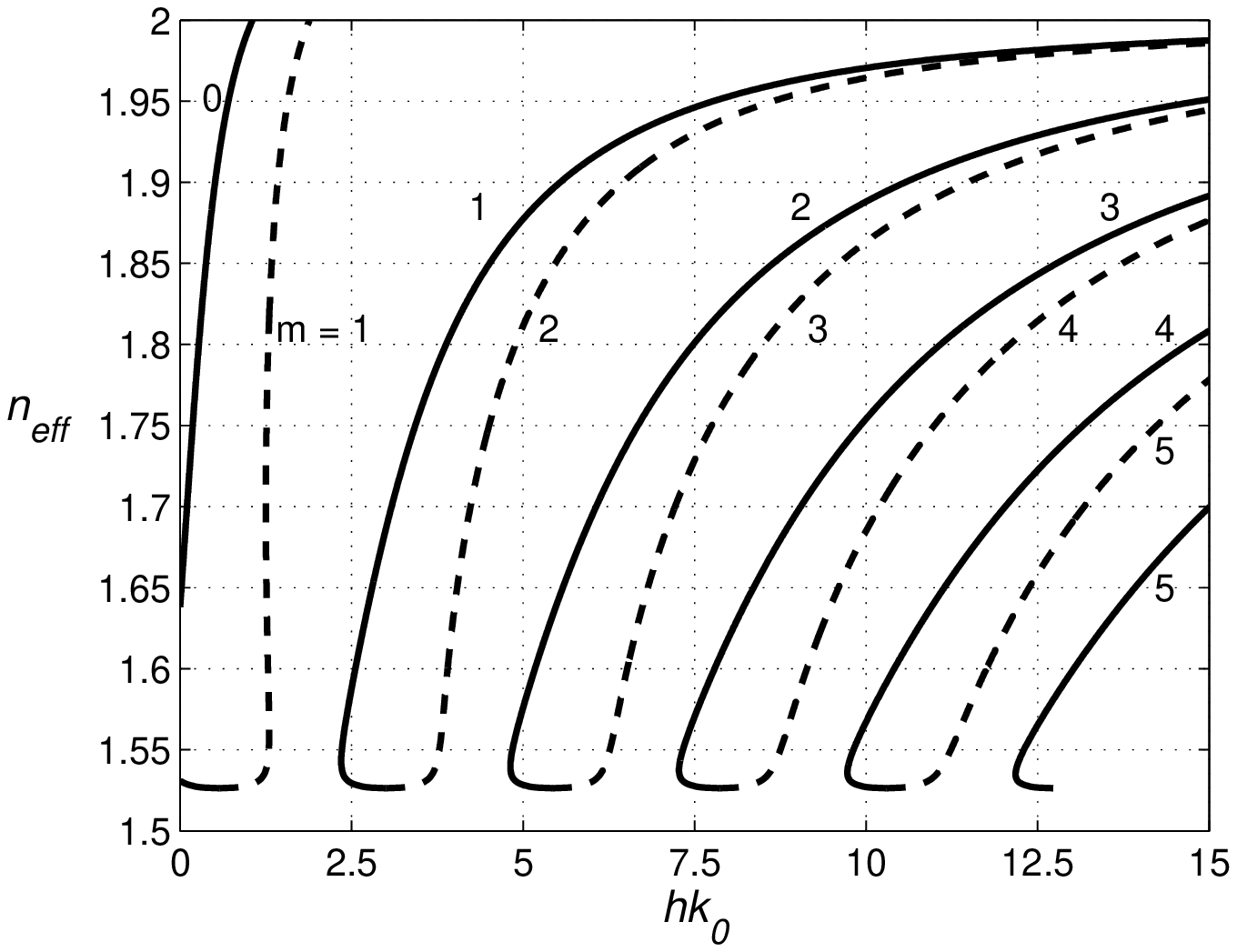}
    \\(a)
\end{minipage}
\hfill
\begin{minipage}[h]{.49\linewidth}
    \centering
    \includegraphics[width=\linewidth]{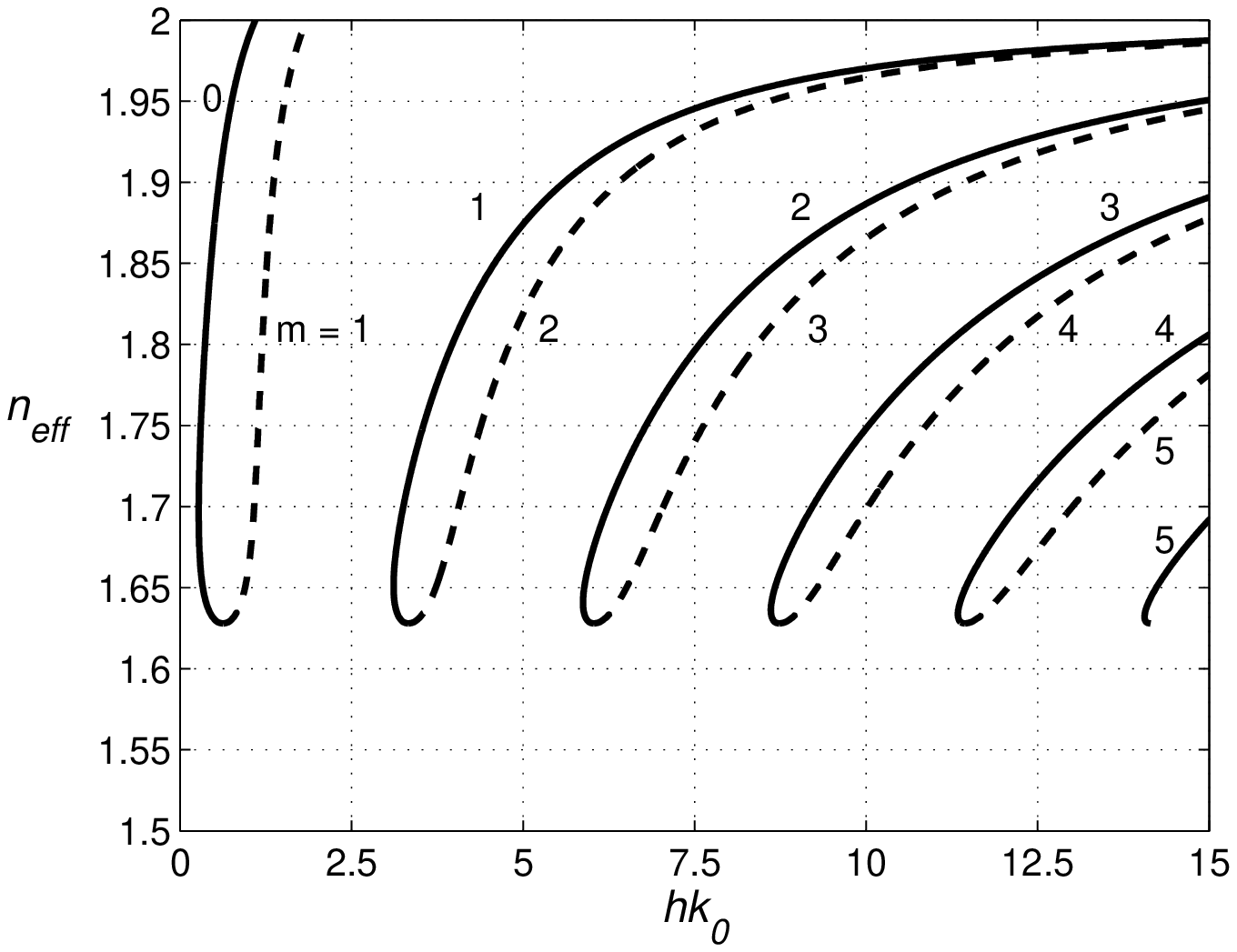}
    \\(b)
\end{minipage}
    \caption{Dispersion curves for different value of $\varepsilon_{nl}$ and mode mark $m$
    at condition $\varepsilon_2<\varepsilon_e$.
    (a) $\varepsilon_{nl}=2.33$, (b) $\varepsilon_{nl}=2.65$.
     }
\label{LM:HLH:8}
\end{figure}

The case of $\varepsilon_2 >\varepsilon_e$ is considered.
Fig.\ref{LM:HLH:8} shows the dispersion curves for different
$\varepsilon_{nl}$ at $\varepsilon_1=2.25$, $\varepsilon_2=4$,
$\varepsilon_e=4.5$, $\varepsilon_0=-3.85$. Mode mark $m$ ranges
from 0 to 5. Solid (dashed) curves correspond to $\nu>0$ ($\nu<0$).
The cut-off frequencies $V_{cm}^{\pm}$ can be defined only for the
curve with marker $m=0$ from branch with $\nu>0$ and the curve
marked $m=1$ from branch with $\nu <0$. The transition points exist
for all curves.

\section{Power fluxes in the nonlinear hyperbolic waveguide}

\noindent The Poynting vector defines density of the radiation
energy flux and direction of wave's energy propagation. It is
instructive to consider an averaged projection of the Poynting
vector along the $OZ$ axis:
\begin{equation}\label{eq:Point:TM}
\langle S_z \rangle =\frac{c}{16\pi}(E_x^*H_y+E_xH_y^*).
\end{equation}
The relation (\ref{eq:Point:TM}) can be rewritten with taking into
account the Maxwell equation for TM wave and the power can be written as
\begin{equation}\label{eq:Point}
P_z =
\frac{c}{8\pi}\int_{-\infty}^{+\infty}\frac{n_{eff}}{\varepsilon_e(x)}|\tilde{H}_y(x)|^2
dx.
\end{equation}

Tangential components of the magnetic field distributions $\tilde{H}_y$
are represented by the following expressions:
\begin{eqnarray}
  x<0   &~& \tilde{H}^{(1)}_y(x) = \frac{ik_0\varepsilon_1\tilde{A}_0\cosh(px_0)}{p}\frac{\tanh[p(x-x_0)]}{\cosh[p(x-x_0)]},  \nonumber\\
  0\leq x\leq h &~& \tilde{H}^{(2)}_y(x) = -\frac{ik_0\varepsilon_2\tilde{A}_0}{\kappa} \left(\sin(\kappa x) + \nu\frac{\varepsilon_1\kappa}
  {\varepsilon_2p}\cos(\kappa x)\right),\label{eq:LM:nonlin:18:H}\\
  x>h   &~& \tilde{H}^{(3)}_y(x) = \frac{ik_0\varepsilon_o\tilde{A}_0}{q}\left(\cos(\kappa h) - \nu\frac{\varepsilon_1\kappa}
  {\varepsilon_2p}\sin(\kappa h)\right)e^{-q(x-h)}.\nonumber
\end{eqnarray}

Total power flux $P_z$ is equal to sum of the power flux in the
substrate $P_z^{(1)}$, the power flux in the dielectric core
$P_z^{(2)}$ and the power flux in the hyperbolic material covering
the dielectric layer $P_z^{(3)}$. These values are calculating
as
$$
P_z^{(1)} = \frac{c n_{eff}}{8\pi
\varepsilon_1}\int_{-\infty}^{0}|\tilde{H}^{(1)}_y(x)|^2 dx, \quad
P_z^{(2)} = \frac{cn_{eff}}{8\pi
\varepsilon_2}\int_{0}^{h}|\tilde{H}^{(2)}_y(x)|^2 dx,\quad
P_z^{(3)} = \frac{cn_{eff}}{8\pi
\varepsilon_e}\int_{h}^{+\infty}|\tilde{H}^{(3)}_y(x)|^2dx.
$$

Using the equations (\ref{eq:LM:nonlin:H:18}) these partial values
of the power flux can be determined:
$$
P_z^{(1)} =
\frac{c}{24\pi}\frac{k_0^2n_{eff}\varepsilon_1}{p^3}A_0^2\left(
\cosh^2(px_0)-\nu \sinh^2(px_0)\right),
$$           
$$
    P_z^{(2)} =
\frac{c}{16\pi}\frac{k_0^2n_{eff}\varepsilon_2}{\kappa^3}A_0^2
\left[\kappa h\left(1+\frac{\nu^2}{\xi_p^2}\right) -
\frac{1}{2}\left(1-\frac{\nu^2}{\xi_p^2}\right)\sin(2\kappa h)+\frac{2\nu
\sin^2(\kappa h)}{\xi_p}\right],
$$           
$$
P_z^{(3)} =
\frac{c}{16\pi}\frac{k_0^2n_{eff}\varepsilon_o^2}{\varepsilon_e
q^3}A_0^2\left(\cos(\kappa h)-\frac{\nu\sin(\kappa h)}{\xi_p}
\right)^2,
$$
where $\xi_p=\varepsilon_2p/(\varepsilon_1\kappa).$

\begin{figure}[h]
\begin{minipage}[h]{.49\linewidth}
    \centering
    \includegraphics[width=\linewidth]{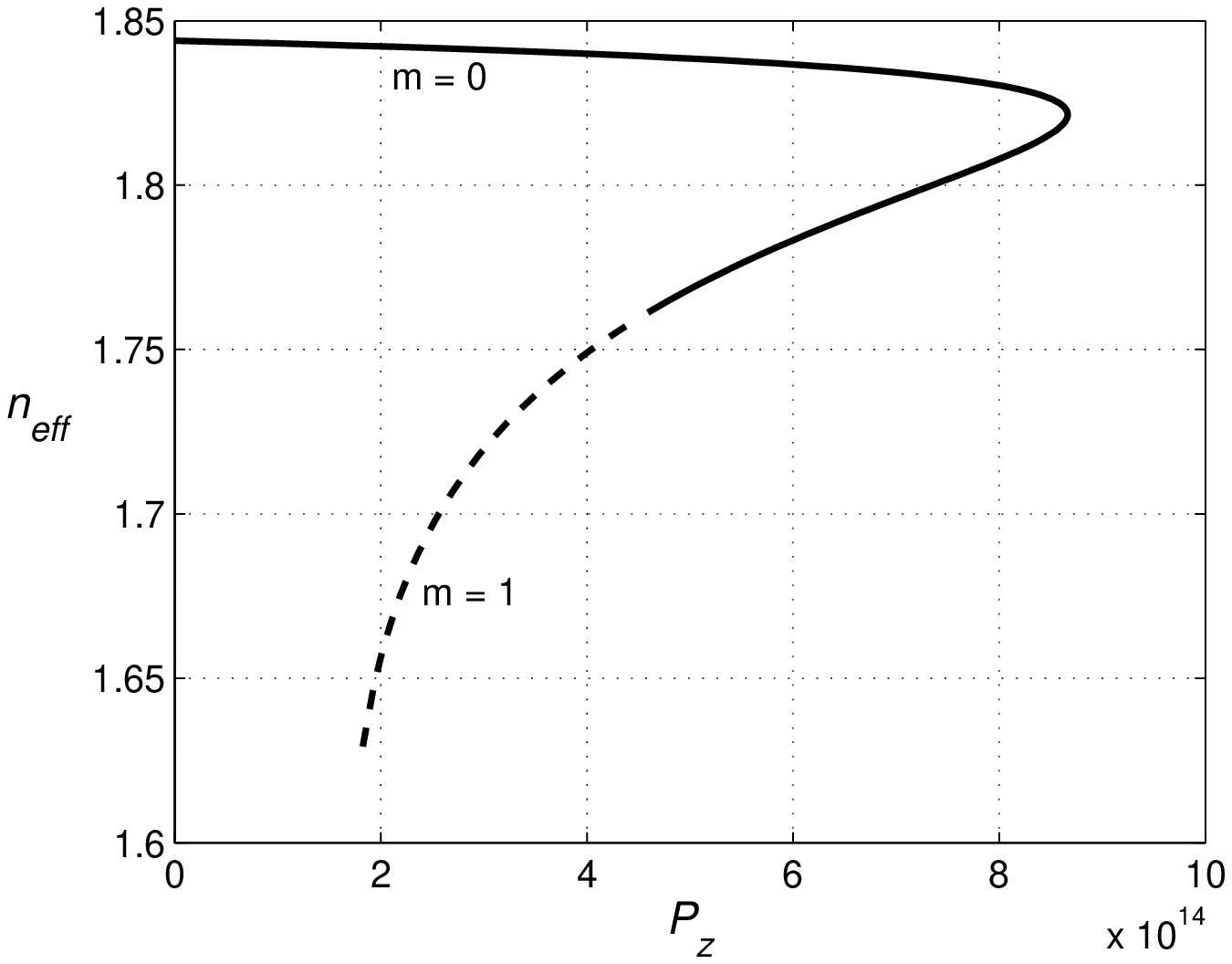}
    \\(a)
\end{minipage}
\hfill
\begin{minipage}[h]{.49\linewidth}
    \centering
    \includegraphics[width=\linewidth]{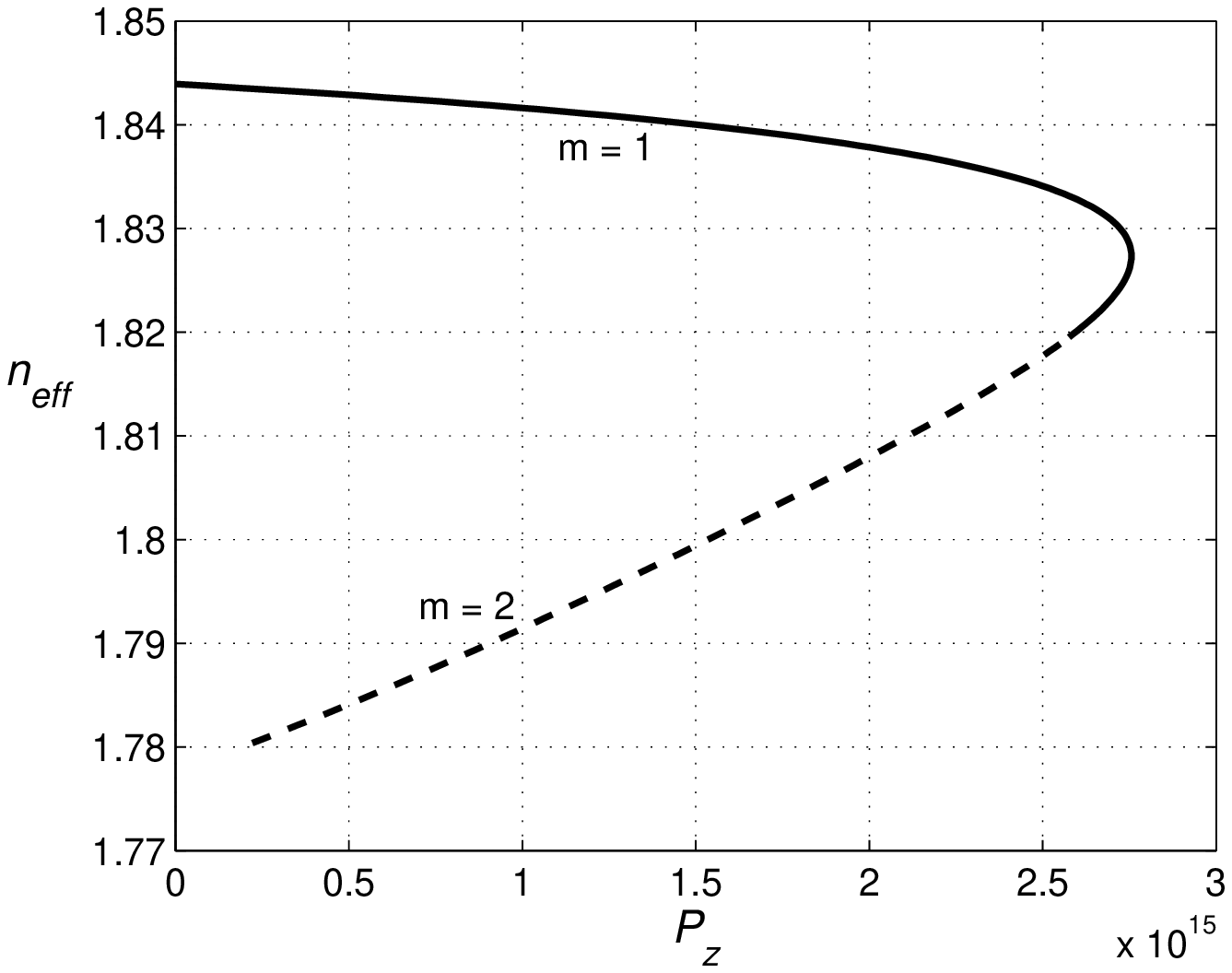}
    \\(b)
\end{minipage}
    \caption{Dependence of the effective index $n_{eff}$ on power. Mode marks
    $m=0$ and $m=1$ are used in ($a$), $m=1$ and $m=2$ are used in ($b$).
     }
\label{LM:HLH:9}
\end{figure}

Due to that parameters $q$, $p$, and $\kappa$ are the functions of
the effective index $n_{eff}$, the total power flux depends on
$n_{eff}$ 
That allows one to consider the different features of the waveguide
as a function of power. Fig.\ref{LM:HLH:9} ($a$) shows the
dependence of the effective index $n_{eff}$ on $P_z$. Solid (dashed)
curves correspond to $\nu>0$ ($\nu<0$). The following parameters of
the waveguide are used: $\varepsilon_1=2.25$, $\varepsilon_2 =4$,
$\varepsilon_e=3.5$, $\varepsilon_0=-3.85$, $\varepsilon_K=10^{-9}$
esu, $k_0h= 1$ at $\lambda $ = 1 $\mu m$. The mode with mark $m=0$
is used in the case of $\nu>0$ and the mode $m=1$ is used for
$\nu<0$. The same is presented in case ($b$), but $k_0h= 5.05$ and
the mode with $m=1$ for $\nu>0$ and the mode with mark $m=2$ for
$\nu<0$ are used for calculations.

There two curves corresponding to $\nu>0$ and $\nu<0$, which are
convergent at value of $P_z$ corresponding to  $A_0^2 =
2/\varepsilon_K(p/k_0)^2$, Fig. \ref{LM:HLH:9}. The curve with
$\nu>0$ presents the dependence of the effective index on total
power flux for a electric field varying with $x$ monotonically in
the substrate.

As can be seen from Fig.\ref{LM:HLH:9} at appropriate thickness of
the waveguide and radiation frequency these modes can be excited
independently of the power value. So, these guided wave modes are
reduced to modes of the linear waveguide (\ref{eq:LM:Lin:6}). To
excite the modes of the $\nu<0$ branch the power must exceed some
threshold value. For all branches the maximum of the power is
limited.

\begin{figure}[h]
    \begin{minipage}[h]{.49\linewidth}
        \centering
        \includegraphics[width=\linewidth]{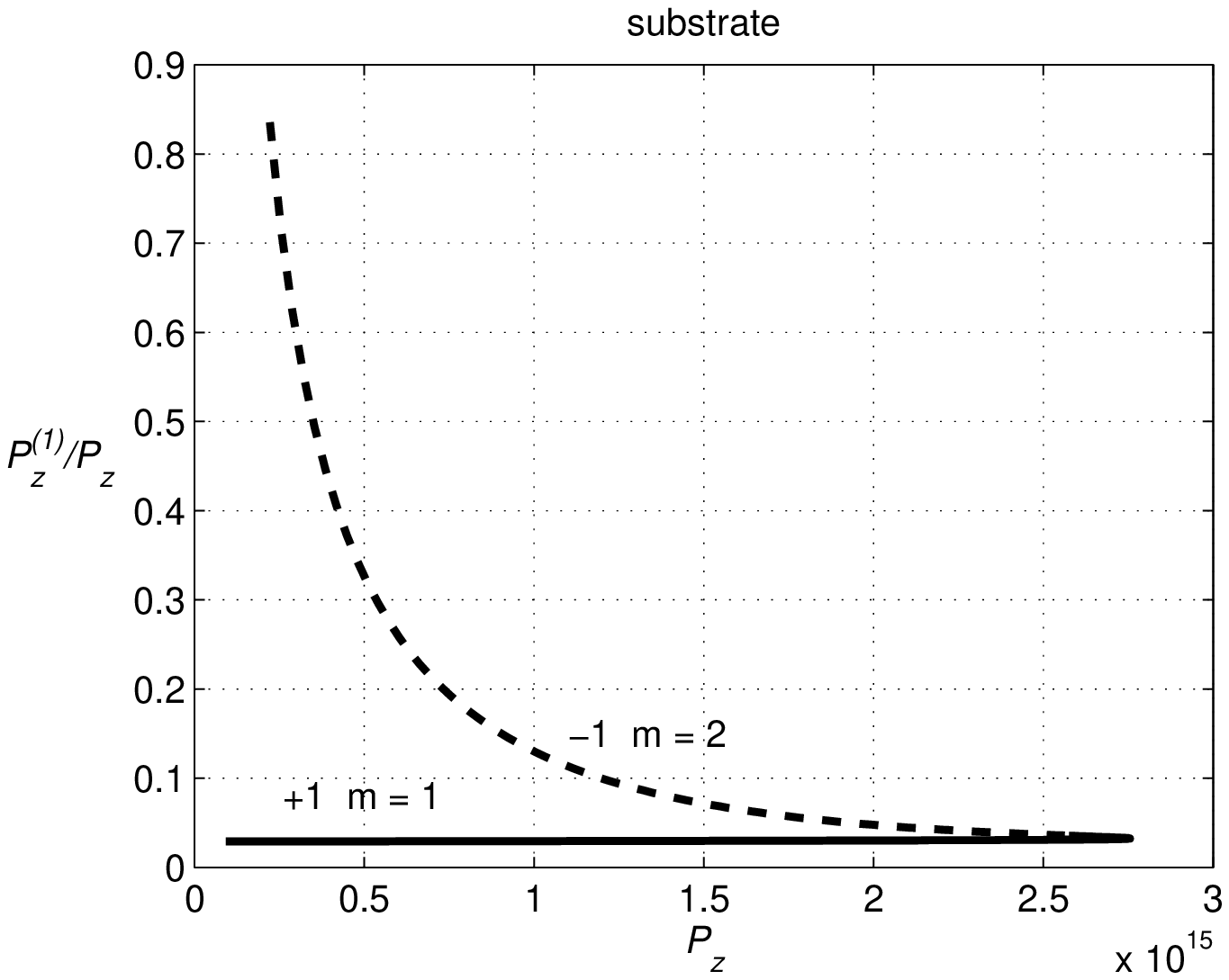}
        \\(a)
    \end{minipage}
    \hfill
    \begin{minipage}[h]{.49\linewidth}
        \centering
        \includegraphics[width=\linewidth]{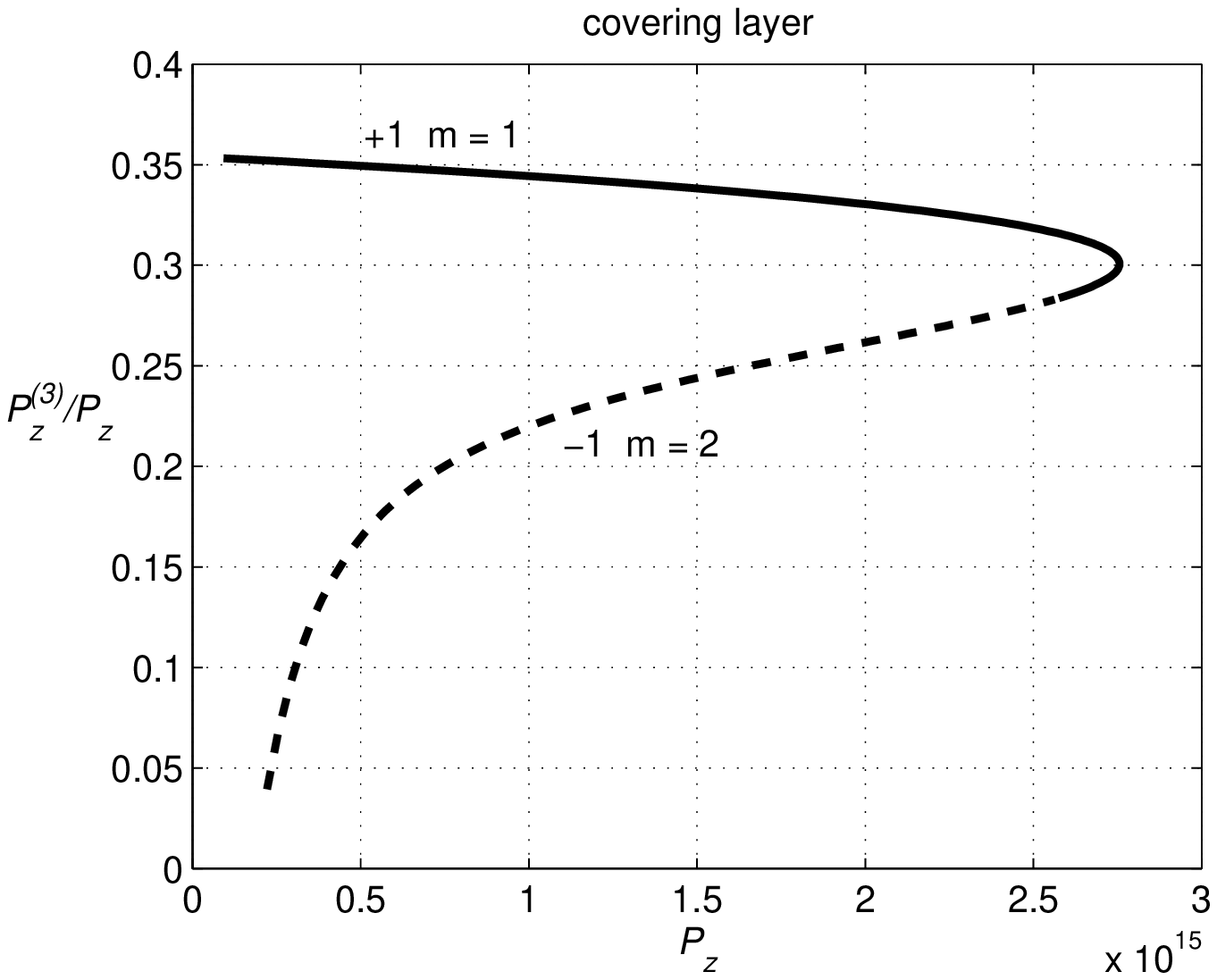}
        \\(c)
    \end{minipage}
    \hfill
    \begin{minipage}[h]{.49\linewidth}
        \centering
        \includegraphics[width=\linewidth]{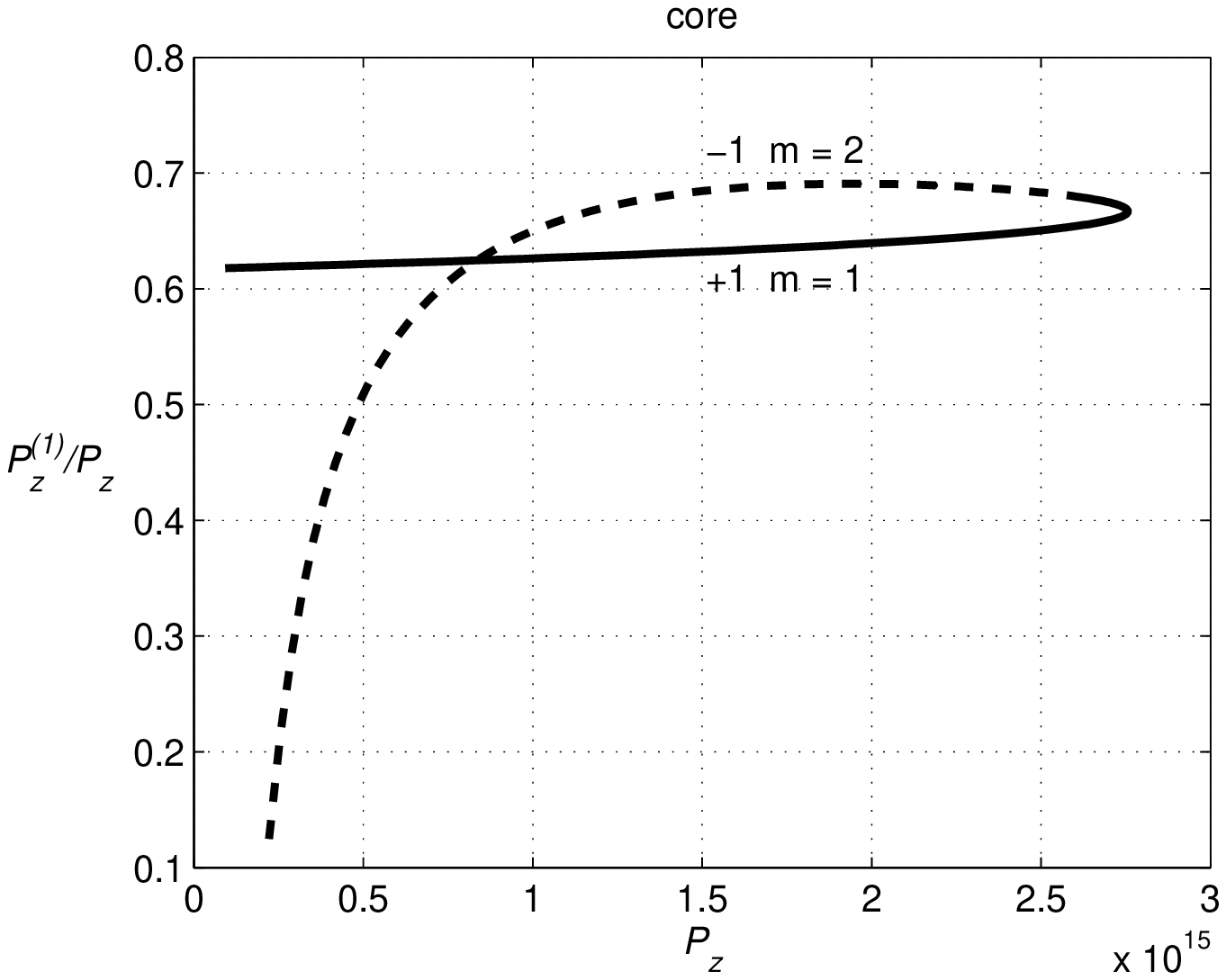}
        \\(b)
    \end{minipage}
    \caption{The power ration $P_z^{(1)}/P_z$ (a), $P_z^{(2)}/P_z$ (b) and
    $P_z^{(3)}/P_z$ as a function of $P_z$. Case $\varepsilon_e < \varepsilon_2$.
    }
\label{LM:HLH:11}
\end{figure}

The plots in Fig. \ref{LM:HLH:11}) demonstrate the proportions of
the power, which are localized in a substrate $P_z^{(1)}/P_z$, in a
core $P_z^{(2)}/P_z$, and in a hyperbolic cover $P_z^{(3)}/P_z$ at
$m=1$ and $m=2$ in case under consideration, $\varepsilon_e <
\varepsilon_2$.

It follows that the radiation of mode with $\nu>0, m=1$ is localized
principally in core and cover. This situation is little affected
with increasing of the transferred power  $P_z$. Oppositely,
proportions of the power transported by mode with $\nu <0, m=2$ are
changed over wide range under increasing of total transferred power.
The principal part of the radiation is localized in the nonlinear
substrate, but with increasing the $P_z$ the portion of radiation in
the core grows. And The power in waveguide core for the modes of the
$\nu<0$ branch can be greater then the  transferred power  by the
modes of the $\nu>0$ branch.

For comparison purposes the same proportions are shown for the case
$\varepsilon_e > \varepsilon_2$ in Fig. \ref{LM:HLH:11}. It can be
seen that the proportion of power in the covering layer is
negligible in comparison with case $\varepsilon_e < \varepsilon_2$.
It can be connected with the absence of the additional cut-off
frequencies on the dielectric--hyperbolic material boundary.

\begin{figure}[h]
    \begin{minipage}[h]{.49\linewidth}
        \centering
        \includegraphics[width=\linewidth]{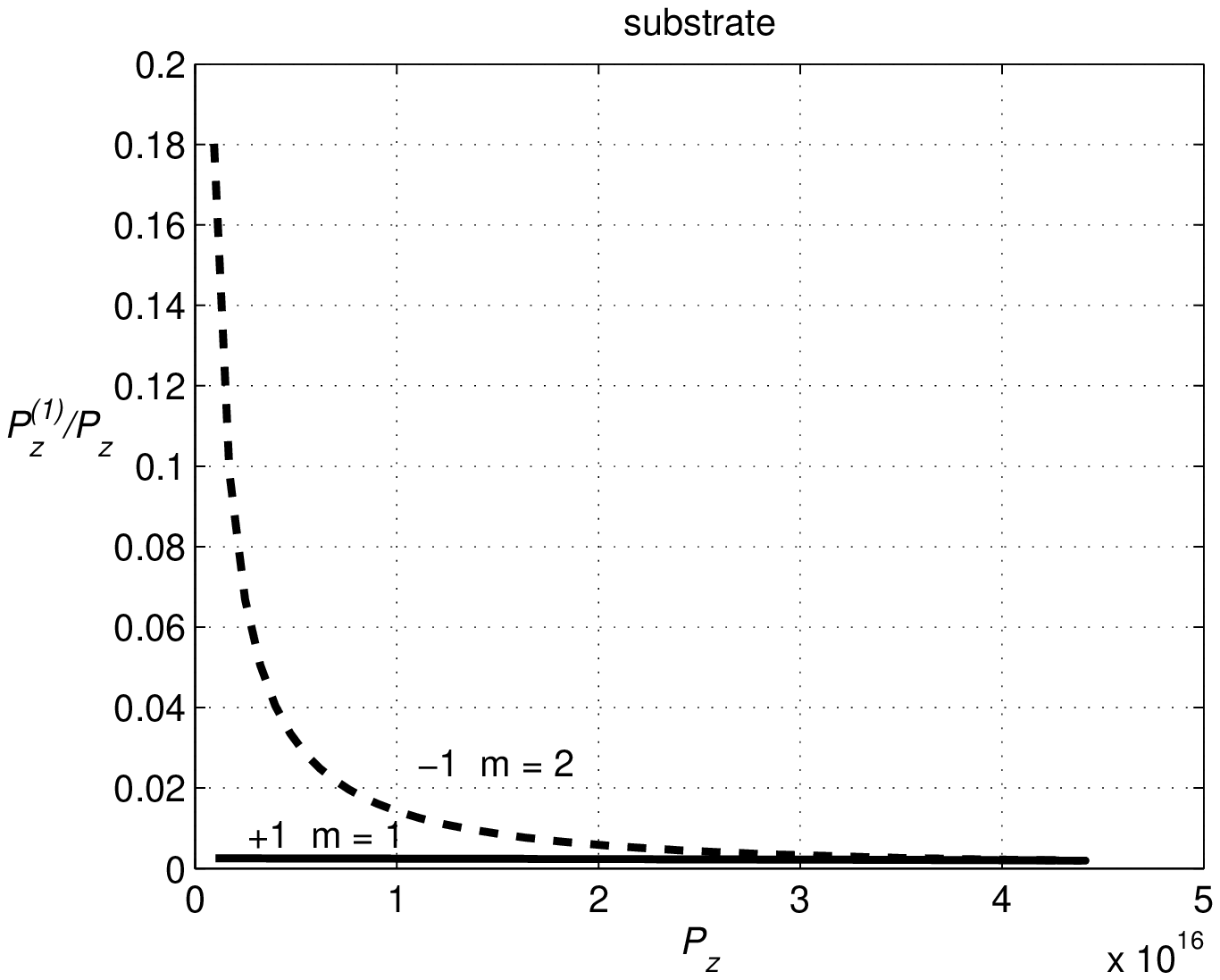}
        \\(a)
    \end{minipage}
    \hfill
    \begin{minipage}[h]{.49\linewidth}
        \centering
        \includegraphics[width=\linewidth]{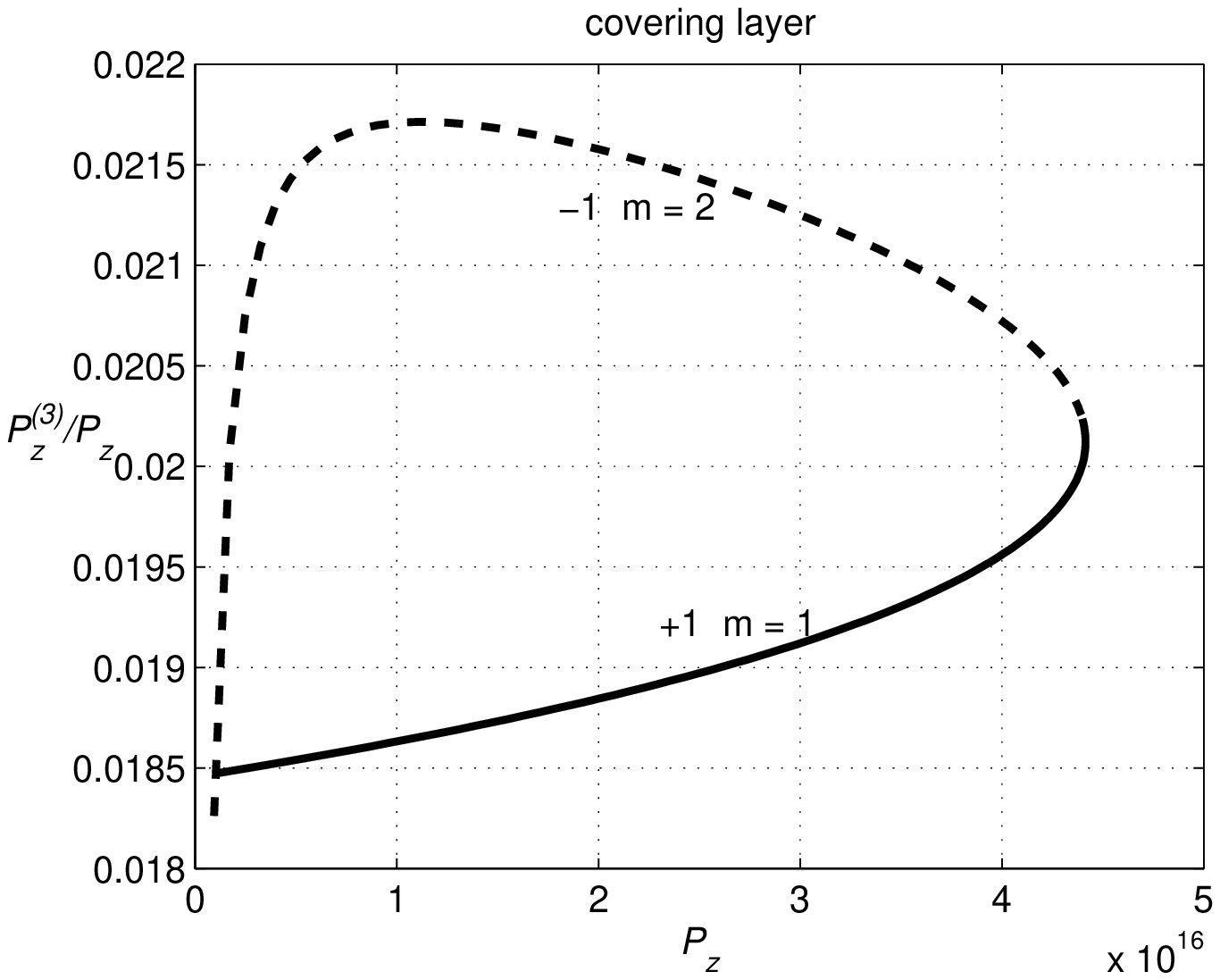}
        \\(c)
    \end{minipage}
    \hfill
    \begin{minipage}[h]{.49\linewidth}
        \centering
        \includegraphics[width=\linewidth]{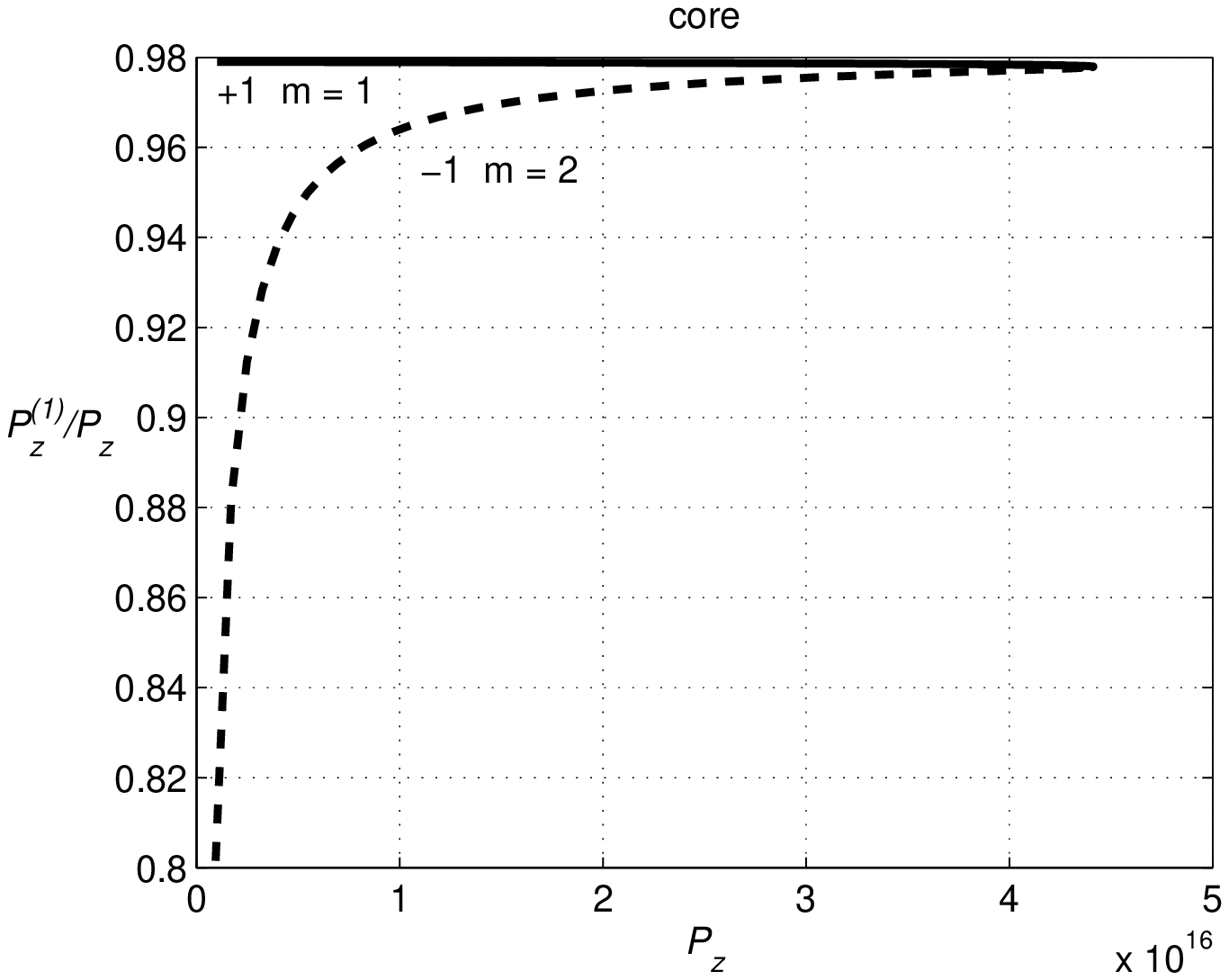}
        \\(b)
    \end{minipage}
    \caption{The power ration $P_z^{(1)}/P_z$ (a), $P_z^{(2)}/P_z$ (b) and
        $P_z^{(3)}/P_z$ as a function of $P_z$. Case $\varepsilon_e > \varepsilon_2$.
    }
\label{LM:HLH:12}
\end{figure}

\section{Conclusion}

In the paper we have considered the TM-polarized linear and
nonlinear guided waves which are guided by a linear dielectric slab
embedded between linear or nonlinear substrate and uniaxial
hyperbolic media. The anisotropy axis is aligned with a unit normal
vector to the interface. In this geometry the TE wave is ordinary
wave and the TM-polarized wave is the extra-ordinary one.
The properties of the cover change the features of the guided TM waves.
Both for a linear and for a nonlinear substrate the electromagnetic
radiation can be confined in the waveguide under conditions $\varepsilon_o <0$, $\varepsilon_e>0$.

The electric and magnetic field transverse distributions and the
dispersion relations are found  in linear and nonlinear cases. In
the linear case under condition that $\varepsilon_e< \varepsilon_2$
any guided wave mode has two cut-off frequencies. One of them
corresponds to mode appearance, another corresponds to mode
disappearance. It follows that there is a region of parameters in
which the only several modes exist. It should be pointed that each
conventional dielectric waveguide mode has only one cut-off
frequency. It is worth noting that this phenomenon is unavailable in
the case of a conventional waveguide. Usually the number of modes
increases with core thickness or radiation frequency, and only
single cut-off frequency exists. If $\varepsilon_e > \varepsilon_2$
then the single cut-off frequency exists for all modes marked $m\geq
1$. However, the fundamental mode with mark $m=0$ has the two
cut-off frequencies. In the case of a conventional asymmetrical
dielectric waveguide the single cut-off frequency is positive
quantity. In the case under consideration one of the cut-off
frequencies of the fundamental mode is to be negative quantity
according to dispersion relation for the hyperbolic waveguide. It is
imposable, thus the real magnitude of effective index is achieved at
the zero film thickness and it is not equal zero. The property of
the guided mode discussed above is the feature of hyperbolic
asymmetrical slab waveguide.

In the nonlinear hyperbolic waveguide the modes can be collected in
two sets. Both the dispersion curves and the modes (i.e., electric
and magnetic field transverse distributions) are labeled by two
numbers. One of them is $\pm 1$ in relation to the sign of the
coordinate of the electric field strength maximum. If the maximum is
localized in the nonlinear substrate then the coordinate $x_0<0$. If
the maximum is localized outside substrate then the coordinate $x_0
>0$, this is the case of virtual maximum. The second item
of the label is integer $m$. It is the mode marker, which is present
in the dispersion relations (\ref{eq:LM:13}) and
(\ref{eq:LM:nonlin:disper:2}). The waveguide modes marked by $(+1,
m)$ are reducible to the linear guiding modes. The modes with label
$(-1, m)$ corresponds with situation where the peak of electric
field is localized in the nonlinear substrate. These modes are
absent in the linear waveguide. To excite these modes the power must
exceed certain threshold value.

As the frequency of radiation or the film thickness changes, the
dispersion curves marked $(+1, m)$ and dispersion curves marked
$(-1, m+1)$ convergent in some point that is referred to as the
transition point \cite{Lederer:85}. For the modes marked $m\geq 1$
these pictures in the case of hyperbolic and the conventional
dielectric waveguide are similar in appearance. The exception is
fundamental mode with $m=0$. The transition point is absent for this
mode.

As in the case of linear case, the nonlinear hyperbolic waveguide
under condition $\varepsilon_e < \varepsilon_2$ is characterized by
the additional cut-off frequencies. Thus the number of modes is
always limited.

{\bf Acknowledgments}

We are grateful to Prof. I. Gabitov and Dr. C. Bayun for
enlightening discussions. This investigation is funded by Russian
Science Foundation (project 14-22-00098).

\end{document}